\documentclass[twoside,twocolumn,9pt]{article}
\usepackage{extsizes}
\usepackage[super,sort&compress,comma]{natbib} 
\usepackage[version=3]{mhchem}
\usepackage[left=1.5cm, right=1.5cm, top=1.785cm, bottom=2.0cm]{geometry}
\usepackage{balance}
\usepackage{mathptmx}
\usepackage{sectsty}
\usepackage{graphicx} 
\usepackage{lastpage}
\usepackage[format=plain,justification=justified,singlelinecheck=false,font={stretch=1.125,small,sf},labelfont=bf,labelsep=space]{caption}
\usepackage{float}
\usepackage{fancyhdr}
\usepackage{fnpos}
\usepackage[english]{babel}
\addto{\captionsenglish}{%
  
}
\usepackage{array}
\usepackage{droidsans}
\usepackage{charter}
\usepackage[T1]{fontenc}
\usepackage[usenames,dvipsnames]{xcolor}
\usepackage{setspace}
\usepackage[compact]{titlesec}
\usepackage{hyperref}

\usepackage{epstopdf}
\usepackage{mathtools, cuted}
\usepackage{lipsum, color}
\definecolor{cream}{RGB}{222,217,201}

\begin{document}

\pagestyle{fancy}
\thispagestyle{plain}
\fancypagestyle{plain}{
\renewcommand{\headrulewidth}{0pt}
}

\makeFNbottom
\makeatletter
\renewcommand\LARGE{\@setfontsize\LARGE{15pt}{17}}
\renewcommand\Large{\@setfontsize\Large{12pt}{14}}
\renewcommand\large{\@setfontsize\large{10pt}{12}}
\renewcommand\footnotesize{\@setfontsize\footnotesize{7pt}{10}}
\makeatother

\renewcommand{\thefootnote}{\fnsymbol{footnote}}
\renewcommand\footnoterule{\vspace*{1pt}%
\color{cream}\hrule width 3.5in height 0.4pt \color{black}\vspace*{5pt}} 
\setcounter{secnumdepth}{5}

\makeatletter 
\renewcommand\@biblabel[1]{#1}            
\renewcommand\@makefntext[1]%
{\noindent\makebox[0pt][r]{\@thefnmark\,}#1}
\makeatother 
\renewcommand{\figurename}{\small{Fig.}~}
\sectionfont{\sffamily\Large}
\subsectionfont{\normalsize}
\subsubsectionfont{\bf}
\setstretch{1.125} 
\setlength{\skip\footins}{0.8cm}
\setlength{\footnotesep}{0.25cm}
\setlength{\jot}{10pt}
\titlespacing*{\section}{0pt}{4pt}{4pt}
\titlespacing*{\subsection}{0pt}{15pt}{1pt}

\fancyfoot{}
\fancyfoot[RO]{\footnotesize{\sffamily{1--\pageref{LastPage} ~\textbar  \hspace{2pt}\thepage}}}
\fancyfoot[LE]{\footnotesize{\sffamily{\thepage~\textbar\hspace{3.45cm} 1--\pageref{LastPage}}}}
\fancyhead{}
\renewcommand{\headrulewidth}{0pt} 
\renewcommand{\footrulewidth}{0pt}
\setlength{\arrayrulewidth}{1pt}
\setlength{\columnsep}{6.5mm}
\setlength\bibsep{1pt}

\makeatletter 
\newlength{\figrulesep} 
\setlength{\figrulesep}{0.5\textfloatsep} 

\newcommand{\topfigrule}{\vspace*{-1pt}%
\noindent{\color{cream}\rule[-\figrulesep]{\columnwidth}{1.5pt}} }

\newcommand{\botfigrule}{\vspace*{-2pt}%
\noindent{\color{cream}\rule[\figrulesep]{\columnwidth}{1.5pt}} }

\newcommand{\dblfigrule}{\vspace*{-1pt}%
\noindent{\color{cream}\rule[-\figrulesep]{\textwidth}{1.5pt}} }

\makeatother

\twocolumn[
  \begin{@twocolumnfalse}
\vspace{3cm}
\sffamily
\begin{tabular}{m{4.5cm} p{13.5cm} }

& \noindent\LARGE{\textbf{Elastic deformations of loaded core-shell systems} 
} \\
\vspace{0.3cm} & \vspace{0.3cm} \\

 & \noindent\large{Jannis Kolker$^{\ast}$\textit{$^{a}$}, Lukas Fischer$^{\ddag}$\textit{$^{b}$}, Andreas M. Menzel\textit{$^{b}$} and Hartmut L\"owen\textit{$^{a}$}} \\

& \noindent\normalsize{{Macroscopic elastic core-shell systems can be generated as toy models to be deformed and haptically studied by hand. On the mesoscale, colloidal core-shell particles and microgels are fabricated and investigated by different types of microscopy. We analyse, using linear elasticity theory, the response of spherical core-shell systems under the influence of a line density of force that is oriented radially and acts along the equator of the outer surface. Interestingly, deformational coupling of the shell to the core can determine the resulting overall appearance in response to the forces. We address various combinations of radii, stiffness, and Poisson ratio of core and shell and illustrate the resulting deformations. Macroscopically, the situation could be realized by wrapping a cord around the equator of a macroscopic model system and pulling it tight. On the mesoscale, colloidal microgel particles symmetrically confined to the interface between two immiscible fluids are pulled radially outward by surface tension.}
} \\

\end{tabular}

 \end{@twocolumnfalse} \vspace{0.6cm}

  ]

\renewcommand*\rmdefault{bch}\normalfont\upshape
\rmfamily
\section*{}
\vspace{-1cm}


\footnotetext{\textit{$^{a}$~
Institut f\"ur Theoretische Physik II, Heinrich-Heine-Universit\"at
D\"usseldorf, Universit\"atsstra\ss e 1, D-40225 D\"usseldorf, Germany.}}
\footnotetext{\textit{$^{b}$~Institut f\"ur Physik, Otto-von-Guericke-Universit\"at Magdeburg, Universit\"atsplatz 2, D-39106 Magdeburg, Germany.}}
\pagebreak
\section{Introduction}
Solid sphere-like core-shell systems containing an inner part, the core, of elastic properties different from a surrounding outer part, the shell, are encountered in various {contexts} on various length scales. 
On large macroscopic scales, many stars, planets and moons can be approximated by a core and a shell of different elasticity~\cite{jacobs2011earth}. 
Jelly sweets covered by a solid layer represent a popular example of not only mechanical or haptic but also culinary {experience}. Conversely, on the mesoscopic colloidal scale and even down to the nanoscale, there are numerous soft matter systems involving core-shell particles. These can be prepared in various ways~\cite{caruso2001nanoengineering,schartl2010current} 
 as spherical colloidal particles with a polymer coating~\cite{pusey1989liquids,lekkerkerker2011depletion,royall2013search}
, as micelles~\cite{forster2004polyelectrolyte}
 or as polymer networks with different crosslinking degrees in the inner and outer part~\cite{rey2020poly,rey2017interfacial,plamper2017functional}. 
Their controlled fabrication is not only pivotal for applications (such as microreactors~\cite{yang2008hollow,liu2019patchy}
, targeted drug delivery~\cite{kataoka2001block,bonacucina2009colloidal} 
 or smart elastic materials~\cite{motornov2010stimuli,forster2002self}). 
They also serve as model systems to tailor effective repulsive square-shoulder potentials~\cite{heyes1992square,bolhuis1997isostructural,denton1997isostructural,jagla1998phase,malescio2003stripe,pauschenwein2008zero,yuste2011structure,norizoe2012particle,pattabhiraman2015stability,gabrielse2017low,somerville2020pattern} 
  and to understand fundamental questions of statistical mechanics such as freezing and glass formation~\cite{pusey1989liquids,ivlev2012complex,gasser2009crystallization,karg2019nanogels}.

Our focus in this work is laid on the coupled elastic deformation of inner and outer part, that is core and shell, respectively. We address spherical elastic systems {when} exposed to a force line density along the equatorial circumference of the shell. This setup is {motivated} by the elasticity problem underlying colloidal core-shell {microgel particles} that are adsorbed to the interface between two immiscible fluids. At their common contact line, the two fluids pull on the shell of the microgel particle approximately in a radially outward direction in a symmetric setup~\cite{bresme2007nanoparticles,Harrer2019,kolker2021interface}. On macroscopic toy model systems, the force densities may be applied by hand, while on even larger, global scales atmospheric effects may lead to equatorially located line-like force densities on planets. An example is the thin {area} of low atmospheric pressure located around the equator of the earth in the inter-tropical convergence zone.

In this paper, we study the underlying elasticity problem. We present a general continuum theory to compute and predict the shape change of an elastic core-shell system when loaded by an equatorial ring of line force density. Importantly not only the shell deforms, but also the inner core, and the two deformations are coupled to each other by the overall architecture. Through this coupling, the core can influence or {even} determine the type of deformation of the shell, although the load is applied from outside to the shell, not to the core. We analyse the resulting change of shape in {detail}, as a function of the relative size of core and shell, different mechanical stiffness of core and shell, as well as their compressibility. {In particular, we include the possibility of an elastic {\it auxetic\/} response~\cite{huang2016negative,ren2018auxetic,scarpa2004trends,lakes1987foam,chan1997fabrication,caddock1989microporous}. The latter is characterized by a negative Poisson ratio, i.e. when stretched along one axis the system expands along the perpendicular axes. Materials exhibiting corresponding elastic properties have been identified, constructed and analysed~\cite{babaee20133d,jiang20183d,kim2018printing,evans1991auxetic}.}
Our study links to previously investigated geometries, particularly spherical one-component systems~\cite{Style2015} 
  or hollow capsules~\cite{hegemann2018elastic} {as special cases}. 
Moreover, our additional predictions can be verified by experiments on different scales.

\section{\label{II}Theory and Geometry}
Within linear elasticity theory, small deformations of elastic materials are described.
The position $\mathbf{r}$ of a material {element} can be mapped to its position $\mathbf{r}'$ in the deformed state by adding the displacement vector $\mathbf{u}$. The displacement field $\mathbf{u}\left(\mathbf{r}\right)$ satisfies the homogeneous Navier-Cauchy equations~\cite{cauchy1828exercices}
\begin{align}
{
(1-2\nu)\nabla^2\mathbf{u}\left(\mathbf{r}\right)+\nabla(\nabla \cdot \mathbf{u}\left(\mathbf{r}\right))=\mathbf{0}
}
\label{eqn:navier-cauchy}
\end{align}
with $-1<\nu\leq1/2$ denoting the Poisson ratio of the elastic substance {in three-dimensional situations}~\cite{LANDAU1986}.
Materials of $\nu=1/2$ are incompressible, while those of negative Poisson ratio are referred to as {\it auxetic\/} materials~\cite{LANDAU1986}. The latter, when stretched along a certain axis, expand to the lateral directions (instead of lateral contraction).
We ignore any force acting on the bulk,
for example gravity. Consequently in the bulk, the right-hand side of Eq.~(\ref{eqn:navier-cauchy}) is set equal to zero.

Furthermore, linear elasticity theory for isotropic materials dictates the stress-strain relation~\cite{LANDAU1986}
\begin{align}
{
\frac{E}{1+\nu}\left(\underline{\boldsymbol{\epsilon}}(\mathbf{r})+\frac{\nu}{1-2\nu}\text{Tr}\left(\underline{\boldsymbol{\epsilon}}(\mathbf{r})\right)\underline{\mathbf{I}}\right)=\underline{\boldsymbol{\sigma}}(\mathbf{r}).
}
\label{eqn:stress_strain}
\end{align}
Eq.~(\ref{eqn:stress_strain}) describes the relationship between the strain tensor
$\underline{\boldsymbol{\epsilon}}\left(\mathbf{r}\right)=\left(\nabla\mathbf{u}\left(\mathbf{r}\right)+\left(\nabla\mathbf{u}\left(\mathbf{r}\right)\right)^{\text{T}}\right)/2$
 as the symmetrized gradient of the displacement field $\mathbf{u}\left(\mathbf{r}\right)$
and the {symmetric Cauchy} stress tensor $\underline{\boldsymbol{\sigma}}$ (we mark second-rank tensors and matrices by an underscore). $E$ is the Young modulus of the elastic material
and $\underline{\mathbf{I}}$ is the unit matrix.
The Young modulus $E$ and the Poisson ratio $\nu$ are sufficient to quantify the properties of a homogeneous isotropic elastic material.

The boundary conditions at the surface of the elastic shell are
\begin{align}
\underline{\sigma}\left(\mathbf{r}\right) \cdot \mathbf{n}=\frac{\lambda}{R_\text{s}}\delta\left(\theta-\frac{\pi}{2}\right)\mathbf{n}.
\label{eqn:boundary}
\end{align}
Here, $\mathbf{n}$ describes the normal unit vector of the surface
and $\delta\left(\theta-\frac{\pi}{2}\right)/R_\text{s}$, with $\delta$ the Dirac delta function, sets the location of the line at which {the loading force line density of amplitude $\lambda$ is acting on the core-shell system}.
We use spherical coordinates so that $\theta=\frac{\pi}{2}$ specifies the equator.

\begin{figure}
\centering
\includegraphics[width=0.475\textwidth]{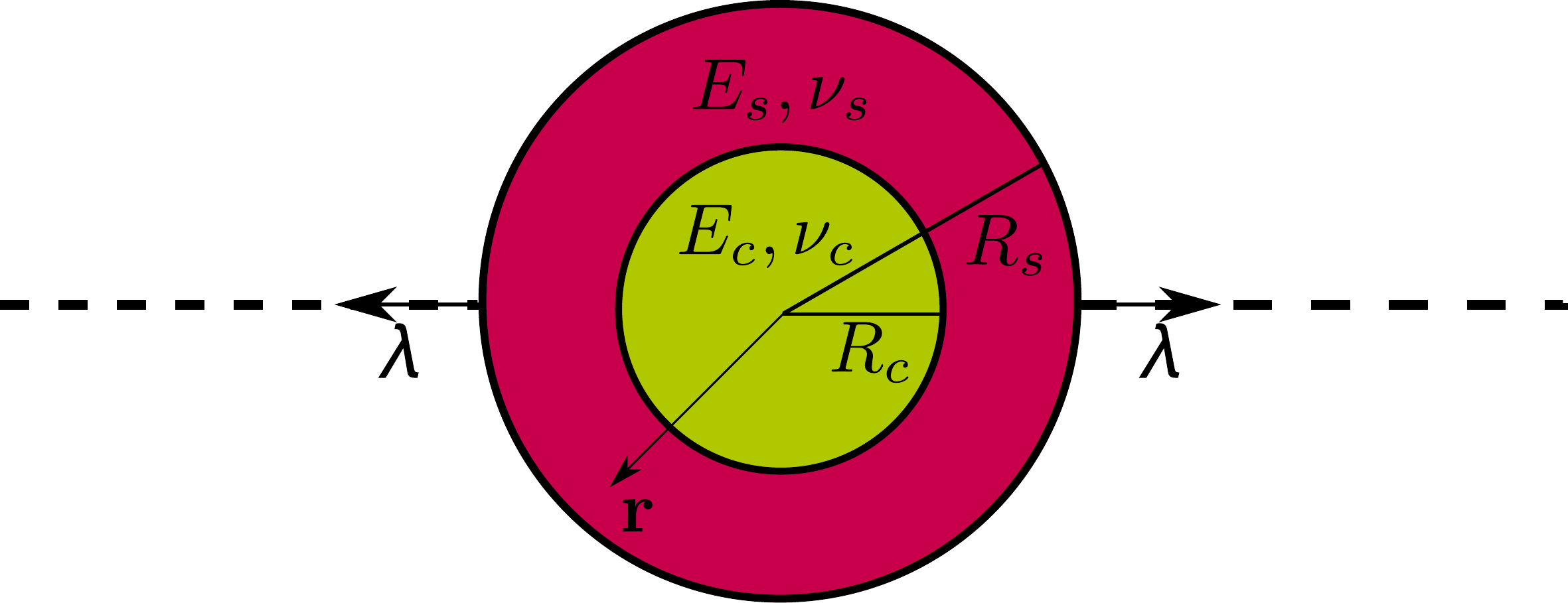}
\caption{{Schematic visualisation of the core-shell system, here still in its initial spherical shape for illustration.
The core (green) is assigned the radius $R_\text{c}$, the Young modulus $E_\text{c}$, and the Poisson ratio $\nu_\text{c}$. The shell (red) is described by the outer radius $R_\text{s}$, the Young modulus $E_\text{s}$, and the Poisson ratio $\nu_\text{s}$. The system is loaded by exposition to a ring of force line density around the equator of the outer sphere of magnitude $\lambda$.
}}
\label{fig:sketch}
\end{figure}

Since we are describing a core-shell material, different elastic properties and radii are attributed to the core and to the shell, see Figure \ref{fig:sketch}. The core (green) is assigned the radius $R_\text{c}$, the Young modulus $E_\text{c}$, and the Poisson ratio $\nu_\text{c}$. The shell (red) is defined by the outer radius $R_\text{s}$, the Young modulus $E_\text{s}$, and the Poisson ratio $\nu_\text{s}$. {In line with Eq.~(\ref{eqn:boundary}), $\lambda>0$ marks the amplitude of a line density of force pointing radially outward along the equator of the outer surface of the shell.}

The system is characterized by the following five dimensionless parameters. First, the ratio $\lambda/E_\text{s}R_\text{s}$ of the loading  force line density on the surface to the Young modulus of the shell describes the relative strength of the load magnitude and is
proportional  to the amplitude of deformation.
The second parameter is the ratio of Young moduli $E_\text{c}/E_\text{s}$ of the core to the shell and in addition, the two dimensionless Poisson ratios $\nu_\text{c}$ and $\nu_\text{s}$ of core and shell, respectively, enter the elasticity theory. The fifth parameter is the size ratio  $R_\text{c}/R_\text{s}$ of the core to the shell.

In spherical coordinates, the position vector $\mathbf{r}$ transforms from the unloaded configuration to the loaded configuration as $\mathbf{r}'=\mathbf{r}+u_r\mathbf{e}_r+u_\theta\mathbf{e}_\theta${, with $u_r$ the radial and $u_\theta$ the polar component of the displacement field. $\mathbf{e}_r$ and $\mathbf{e}_\theta$ denote the radial and polar unit vector, respectively.} Due to {the special} axial symmetry {of the problem}, the azimuthal component of the displacement field, $u_\phi$, is zero. For the homogeneous Navier-Cauchy equations {Eq.~(\ref{eqn:navier-cauchy})} and the stress-strain relation {Eq.~(\ref{eqn:stress_strain}) recast} in spherical coordinates{, where in our case the} azimuthal dependence {vanishes}, see {the Supporting Information (SI)}.

For both core and shell
we solve Eq.~(\ref{eqn:navier-cauchy}) by separation into a
{series expansion of the polar dependence in terms of Legendre polynomials $P_n\left(\cos\theta\right)$ and associated $r$-dependent prefactors $\left(r=\left|\mathbf{r}\right|\right)$}{~\cite{love1927treatise}}.
We distinguish by superscripts c and s the solutions for core and shell, respectively. More precisely, the solutions~\cite{Duan2005,Yi2007,Style2015} of the Navier-Cauchy equations~(\ref{eqn:navier-cauchy}) split into a radial component $u^\text{c}_r\left(\mathbf{r}\right)$ and a polar component $u^\text{c}_\theta\left(\mathbf{r}\right)$ for the core and take the form
\begin{align}
\label{sol1}
u^{\text{c}}_r\left(\mathbf{r}\right)=&\sum_{n=0}^\infty \left(a^{\text{c}}_n(n+1)(-2+n+4\nu_\text{c})r^{n+1}+b^{\text{c}}_n\,nr^{n-1}\right)P_n\left(\cos\theta\right),\\
u^{\text{c}}_\theta\left(\mathbf{r}\right)=&\sum_{n=1}^\infty \left(a^{\text{c}}_n(5+n-4\nu_\text{c})r^{n+1}+b^{\text{c}}_n\,r^{n-1}\right)\frac{d}{d\theta}P_n\left(\cos\theta\right).
\label{sol11}
\end{align}
The solutions for the shell
additionally contain terms inverse in the radial distance from the origin
\begin{align}
\label{sol2}
u_r^{\text{s}}\left(\mathbf{r}\right)=\sum_{n=0}^\infty &\left(a_n^{\text{s}}(n+1)(-2+n+4\nu_\text{s})r^{n+1}+b_n^{\text{s}}\,nr^{n-1}\right.\nonumber\\
&\left.+n(3+n-4\nu_\text{s})c^{\text{s}}_n\,r^{-n}-(n+1)d_n^{\text{s}}\,r^{-(n+2)}\right) P_n\left(\cos\theta\right),\\
u_\theta^{\text{s}}\left(\mathbf{r}\right)=\sum_{n=1}^\infty &\left(a_n^{\text{s}}(5+n-4\nu_\text{s})r^{n+1}+b_n^{\text{s}}\,r^{n-1}\right.\nonumber\\
&\left.-(-4+n+4\nu_\text{s})c_n^{\text{s}}\,r^{-n}+d_n^{\text{s}}\,r^{-(n+2)}\right)\frac{d}{d\theta}P_n\left(\cos\theta\right).
\label{sol21}
\end{align}

As boundary conditions, we use that the
traction vectors at the
interface of core and shell (at radius $R_\text{c}$) must be equal
\begin{align}
\underline{\sigma}^{\text{c}}\left(R_\text{c}\mathbf{e}_r\right)\cdot \mathbf{n} =\underline{\sigma}^{\text{s}}\left(R_\text{c}\mathbf{e}_r\right)\cdot \mathbf{n}.
\end{align}
{Requiring} strict elastic no-slip coupling, also the deformations at the interface must be equal
\begin{align}
\mathbf{u}^{\text{c}}\left(R_\text{c}\mathbf{e}_r\right)=\mathbf{u}^{\text{s}}\left(R_\text{c}\mathbf{e}_r\right).
\end{align}
 Since the Legendre polynomials form a complete orthogonal set, also the Dirac delta function in Eq.~(\ref{eqn:boundary}) can be expanded in Legendre polynomials
\begin{align}
\delta\left(\theta-\frac{\pi}{2}\right)=\sum_{n=0}^\infty\frac{2n+1}{2}P_n\left(\cos\left(\frac{\pi}{2}\right)\right)P_n\left(\cos\theta\right).
\end{align}

Due to the {assumed} mirror symmetry with respect to the equatorial plane, all odd series expansion components in  the core and shell solution in Eqs.~(\ref{sol1})-(\ref{sol21}) vanish. Therefore we can write for the radial displacement 
\begin{align}
u_r^{(i)}\left(\mathbf{r}\right)=u_{r,0}^{(i)}\left(r\right)+u_{r,2}^{(i)}\left(r\right)P_2\left(\cos(\theta)\right)+\dots 
\end{align}
with $i=c$ for the core and $i=s$ for the shell, respectively. Here,
the first component $u^{(i)}_{r,0}(r)$ describes the overall volume change. We note that this term will vanish for $\nu_i \rightarrow 1/2$ {and remains as the only component for $\nu_i \rightarrow -1$}. The second component gives the first correction to a spherical shape.  A positive prefactor $u_{r,2}^{(i)}(r)$ describes a relative prolate deformation  while $u_{r,2}^{(i)}(r)<0$ implies a relative oblate deformation. It is in fact the latter case of an oblate deformation which we expect because the core-shell particle is pulled outwards at the equator $(\lambda >0)$.

The solutions for the displacements of the core and the shell diverge in response to the Dirac delta function acting at the equator on the surface of the shell, see the boundary condition Eq.~(\ref{eqn:boundary}). However, the second components ${u_{r,2}^{(c)}(r)}$ and ${u_{r,2}^{(s)}(r)}$ for the core and the shell are finite even at the surface of the shell. We shall therefore use them as parameters to characterise the anisotropy of the core and the shell shape.

For convenience, we evaluate these second components at the core and shell radii and normalize them  with the corresponding unloaded radii of the core and the shell, respectively. Hence, we use subsequently ${u_{r,2}^{(c)}}/{R_c}\equiv {u_{r,2}^{(c)}(R_c)}/{R_c}$ and ${u_{r,2}^{(s)}}/{R_s}\equiv {u_{r,2}^{(s)}(R_s)}/{R_s}$ as  dimensionless measures for the relative shape anisotropy of the core and the shell.
\section{Results and Discussion}
\subsection{General solution and limiting behaviour}
We first present the solutions for the displacements under the prescribed boundary conditions
by providing the core coefficients of the expansions (\ref{sol1}) and (\ref{sol11})
\begin{align}
a^\text{c}_n=&\frac{\lambda}{E_\text{s}R_\text{s}}\frac{2n+1}{2}P_n\left(0\right)\left(\frac{R_\text{c}}{R_\text{s}}\right)^{-2}R_\text{s}^{-n}\left[\left(\frac{E_\text{c}}{E_\text{s}}\right)\tilde{c}_{01,n}+\tilde{c}_{02,n}\right]\frac{1}{D},\\
b^\text{c}_n=&-\frac{\lambda}{E_\text{s}R_\text{s}}\frac{2n+1}{2}P_n\left(0\right)R_\text{s}^{-(n-2)}\left[\left(\frac{E_\text{c}}{E_\text{s}}\right)\tilde{c}_{03,n}+\tilde{c}_{04,n}\right]\frac{1}{D},
\end{align}
and the shell coefficients of the expansions (\ref{sol2}) and (\ref{sol21})
\begin{align}
a^\text{s}_n=&\frac{\lambda}{E_\text{s}R_\text{s}}\frac{2n+1}{2}P_n\left(0\right)R_\text{s}^{-n}\nonumber\\&
\times\left[\left(\frac{E_\text{c}}{E_\text{s}}\right)^2\tilde{c}_{05,n}+\left(\frac{E_\text{c}}{E_\text{s}}\right)\tilde{c}_{06,n}+\tilde{c}_{07,n}\right]\frac{1}{D},\\
b^\text{s}_n=&-\frac{\lambda}{E_\text{s}R_\text{s}}\frac{2n+1}{2}P_n\left(0\right)R_\text{s}^{-(n-2)}\nonumber\\&
\times\left[\left(\frac{E_\text{c}}{E_\text{s}}\right)^2\tilde{c}_{08,n}+\left(\frac{E_\text{c}}{E_\text{s}}\right)\tilde{c}_{09,n}+\tilde{c}_{10,n}\right]\frac{1}{D},\\
c^\text{s}_n=&\frac{\lambda}{E_\text{s}R_\text{s}}\frac{2n+1}{2}P_n\left(0\right)\left(\frac{R_\text{c}}{R_\text{s}}\right)^{n-1}R_\text{c}^{n}R_\text{s}\nonumber\\&\times
\left[\left(\frac{E_\text{c}}{E_\text{s}}\right)^2\tilde{c}_{11,n}+\left(\frac{E_\text{c}}{E_\text{s}}\right)\tilde{c}_{12,n}+\tilde{c}_{13,n}\right]\frac{1}{D},\\
d^\text{s}_n=&-\frac{\lambda}{E_\text{s}R_\text{s}}\frac{2n+1}{2}P_n\left(0\right)\left(\frac{R_\text{c}}{R_\text{s}}\right)^{n-1}R_\text{c}^{n+2}R_\text{s}\nonumber \\&\times
\left[\left(\frac{E_\text{c}}{E_\text{s}}\right)^2\tilde{c}_{14,n}+\left(\frac{E_\text{c}}{E_\text{s}}\right)\tilde{c}_{15,n}+\tilde{c}_{16,n}\right]\frac{1}{D},
\end{align}
with
\begin{align}
D=
\left(\frac{E_\text{c}}{E_\text{s}}\right)^2\tilde{c}_{17,n}+\frac{E_\text{c}}{E_\text{s}}\tilde{c}_{18,n}+\tilde{c}_{19,n}.
\end{align}
The constants $\tilde{c}_{01,n}$ to $\tilde{c}_{19,n}$ are listed in the SI.
In the absence of a core, i.e. $R_\text{c}\to 0$, or in the absence of the shell, i.e.  $R_\text{c}\to R_\text{s}$, we recover the previous solution for a one-component system
as given in Ref.~\citenum{Style2015}.
Also for the special case of
$E_\text{c}=E_\text{s}$ and $v_\text{c}=v_\text{s}$ of identical core and shell  elasticities, our solution
reduces to that of a one-component system.

\subsection{Relative deformation of the shell and the core}
\begin{figure}
\centering
\includegraphics[width=0.5\textwidth]{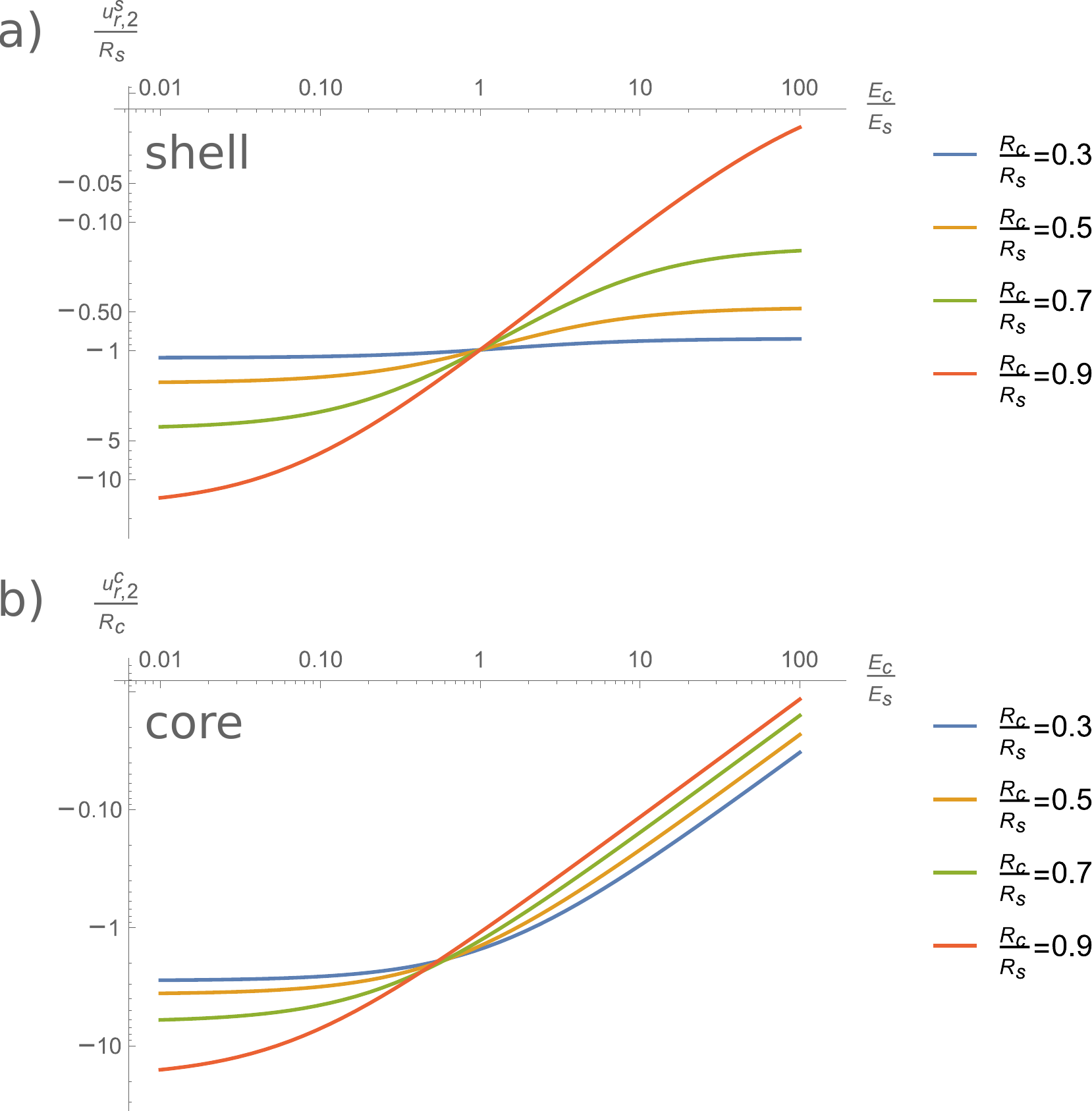}\caption{
Relative
oblate  deformation $u_{r,2}^{(i)}/R_i$
as a function of the ratio of Young moduli $E_\text{c}/E_\text{s}$ at different size ratios $R_\text{c}/R_\text{s}$ for a) the shell ($i=s$) and b) the core ($i=c$) on double logarithmic scale. The further parameters are $\nu_\text{c}=\nu_\text{s}=1/2$ and  $\lambda/(E_\text{s}R_\text{s})=1$.
}
\label{fig:soft_hard_core}
\end{figure}

In the following, the degrees of deformation of the core and the shell
are investigated for volume conserving conditions $\left(\nu_\text{c}=\nu_\text{s} = 1/2\right)$. Figure \ref{fig:soft_hard_core} shows the relative oblate deformation $u_{r,2}^{(i)}/R_i$  for a) the shell ($i=s$) and b) the core ($i=c$) as a function of the ratios of Young moduli $E_\text{c}/E_\text{s}$. Data are given for several size ratios $R_\text{c}/R_\text{s}$ ranging from 0.3 to 0.9.

The first observation is that the coefficient $u_{r,2}^{(i)}/R_i$ is negative, corresponding to a relative oblate deformation. This is a simple consequence of the force load pulling the equator to the outward direction.

Second, the absolute magnitude of deformation decreases with increasing $E_\text{c}/E_\text{s}$ which is the expected trend if the core is getting harder than the shell (at fixed shell elasticity). For $E_\text{c}/E_\text{s}\to 0$ we obtain the special case of a hollow sphere. In this limit, the relative deformation of the core and the shell reaches a finite saturation (note the logarithmic scale in Figure 2). In the opposite limit $E_\text{c}/E_\text{s}\to \infty$ the core gets rigid, which implies that the displacement of the shell stays finite but the displacement of the core tends to zero. We find a common finite slope of -1 for the curves associated with the core for $E_\text{c}/E_\text{s}\to \infty$ in Figure \ref{fig:soft_hard_core}b.

Moreover, in Figure \ref{fig:soft_hard_core}a all curves intersect in the same point at $E_c=E_s$. At this point the two materials are identical and the size ratio becomes irrelevant for the deformation at the shell surface. The curves of Figure \ref{fig:soft_hard_core}b do not exhibit a common intersection point due to our normalization of the relative deformation with $R_c$. For increasing $R_\text{c}/R_\text{s}$, the influence of the core grows and the curves exhibit more sensibility as a function of $E_\text{c}$ for fixed $E_\text{s}$.

\begin{figure}
\centering
\includegraphics[width=0.5\textwidth]{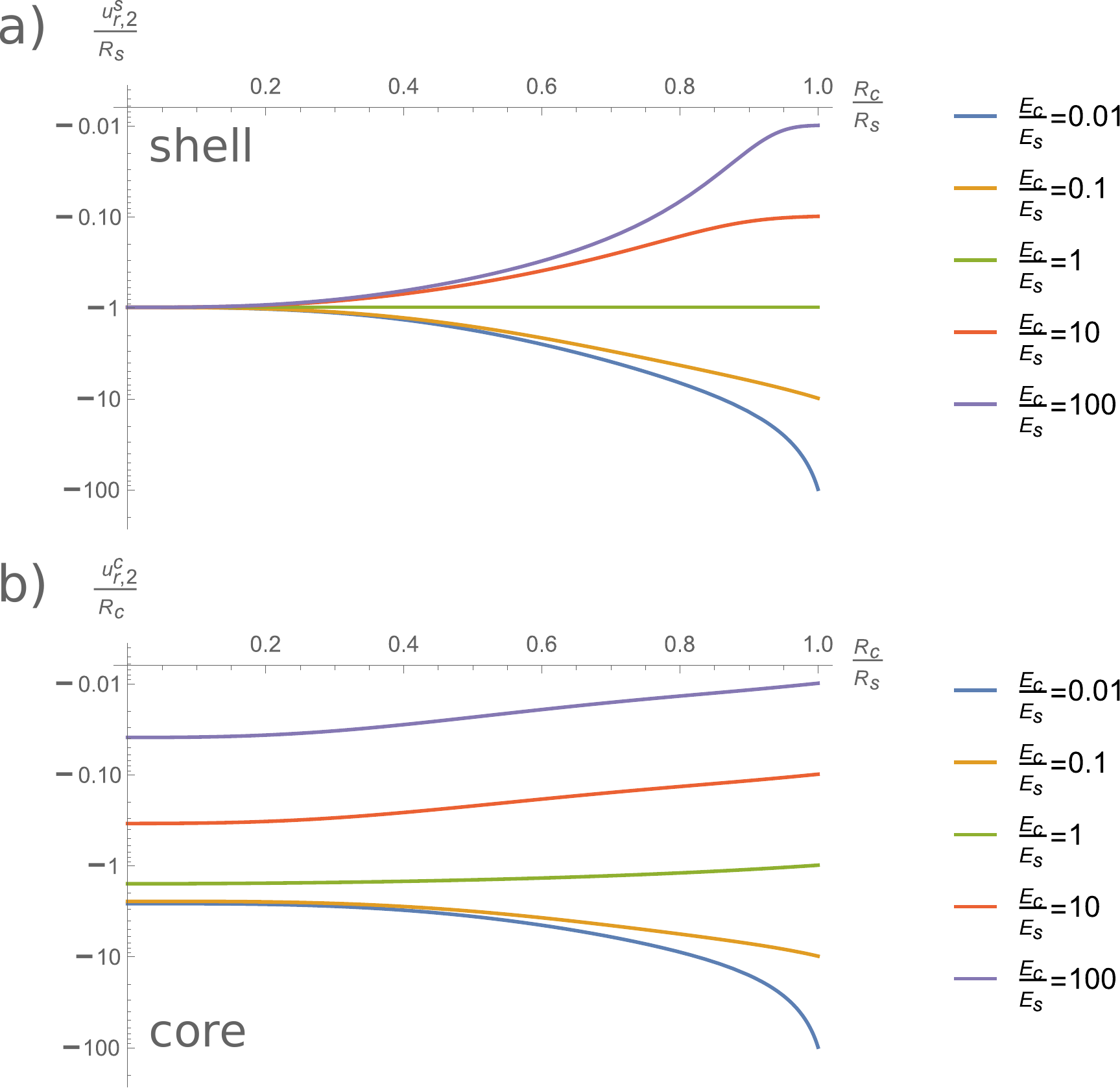}\caption{
Relative
oblate deformation $u_{r,2}^{(i)}/R_i$
as a function of the
size ratio $R_\text{c}/R_\text{s}$ for different ratios of Young moduli $E_\text{c}/E_\text{s}$ for a) the shell ($i=s$) and b) the core ($i=c$) on semi-logarithmic scale. The further parameters are $\nu_\text{c}=\nu_\text{s}=1/2$ and  $\lambda/(E_\text{s}R_\text{s})=1$.
}
\label{fig:deformation_field}
\end{figure}

To complement the picture, Figure \ref{fig:deformation_field} shows the same quantity as in Figure \ref{fig:soft_hard_core}, namely the relative oblate deformation $u_{r,2}^{(i)}/R_i$, but now as a function of the size ratio $R_\text{c}/R_\text{s}$ for a) the shell ($i=s$)  and b) the core ($i=c$). Curves for several ratios of Young moduli $E_\text{c}/E_\text{s}$ are displayed. Again, for $E_\text{c}=E_\text{s}$ (green curves), the resulting effective one-component system features a shell displacement that does not depend on the size ratio of core to shell. Conversely, the plotted core displacement does depend on the size ratio for $E_\text{c}=E_\text{s}$ because it is normalized by the size of the core. The deformation scaled by $R_c$ in the limit of small core size $R_\text{c}\to 0$ (see Figure \ref{fig:deformation_field}b) reaches different limits for different ratios of Young moduli although the core becomes vanishingly small.

\begin{figure*}
\centering
\includegraphics[width=1\textwidth]{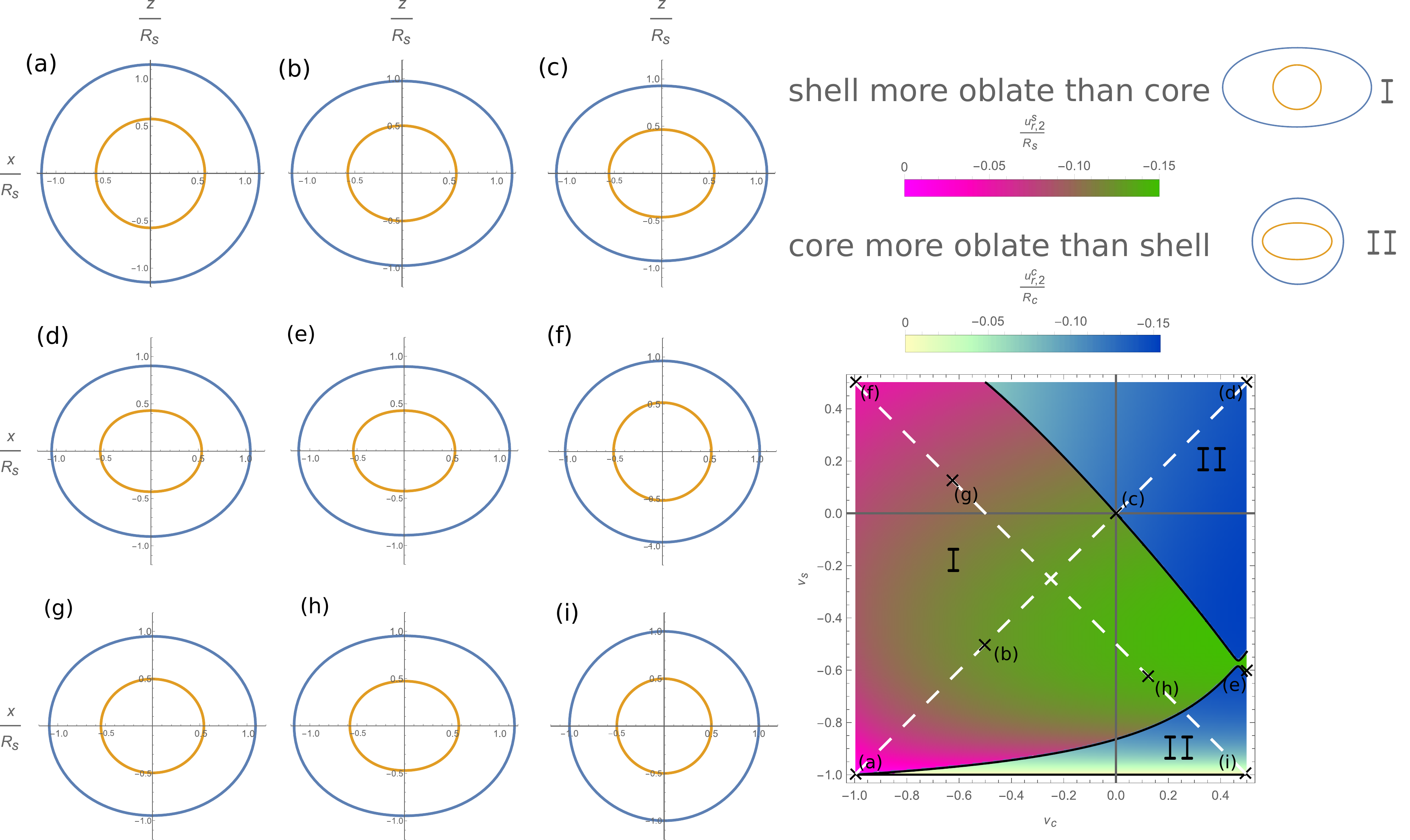}\caption{
Bottom right: State diagram exhibiting two situations I) and II) in the plane spanned by the two Poisson ratios of the core $\nu_\text{c}$ and the shell $\nu_\text{s}$ at fixed  $E_\text{c}=E_\text{s}$,  $R_\text{c}=0.5R_\text{s}$ and  $\lambda/(E_\text{s}R_\text{s})=0.1$. In I), corresponding to the reddish and greenish region, the relative oblate deformation of the shell is larger than that of the core, see schematic representation on the top right. Here we plot in region I of the state diagram {$u^\text{s}_{r,2} / R_\text{s}$} as color-coded on the top right. Conversely, in II), corresponding to the blueish region in the state diagram, the relative oblate deformation of the core is smaller than that of the shell. Here we plot in region II  of the state diagram {$u^\text{c}_{r,2} / R_\text{c}$} as color-coded on the top right. The two states I) and II) are separated by black lines. Furthermore, for nine parameter combinations  indicated for various points (a)-(i) in the state diagram, the corresponding elliptical shapes of core and shell are shown on the left.
}
\label{fig:deformation_poisson}
\end{figure*}
\begin{figure*}
\centering
\includegraphics[width=1.\textwidth]{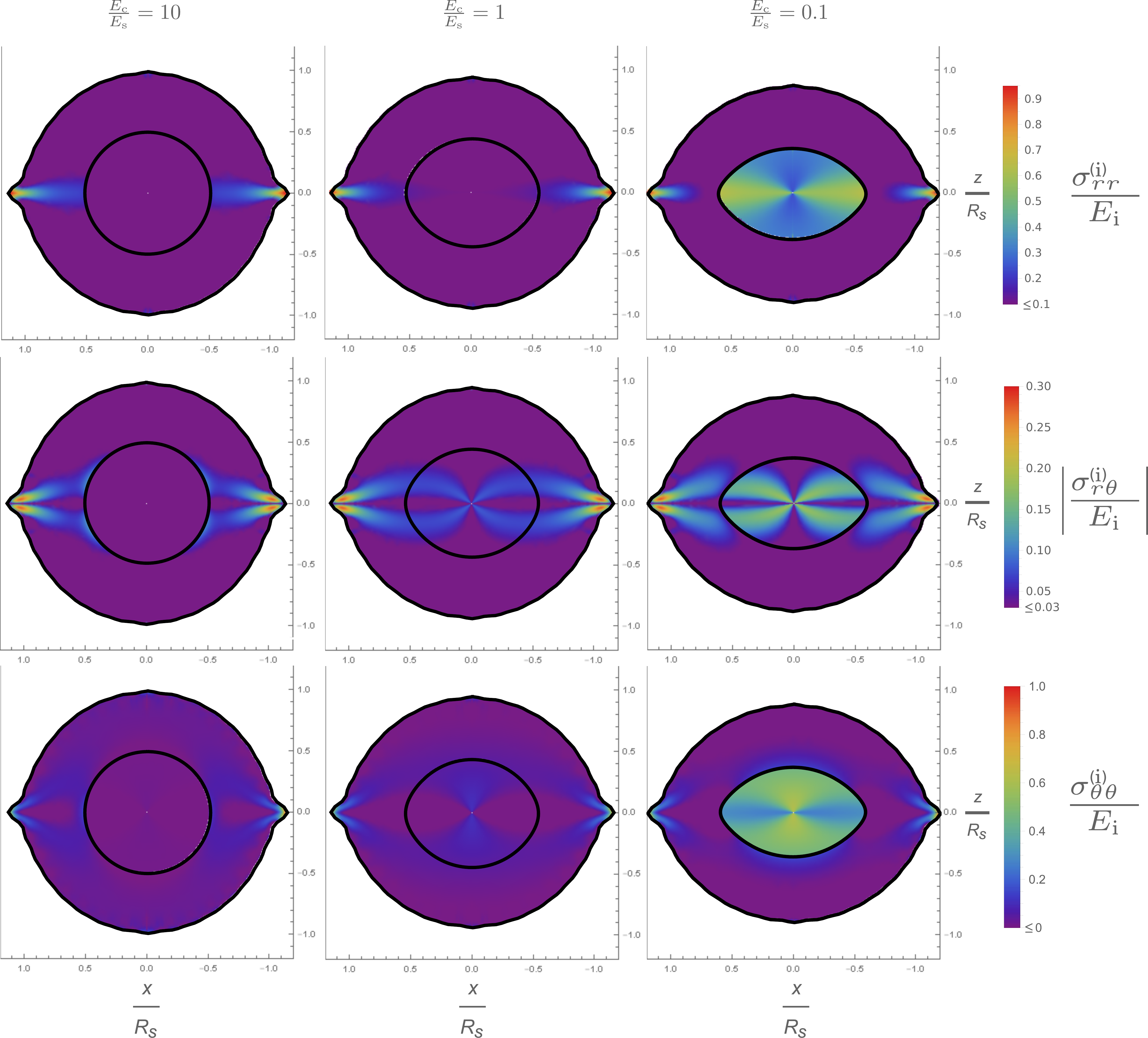}
\caption{
Loaded configurations of the core-shell system at fixed $\nu_\text{c}=\nu_\text{s}=0.4999$, $R_\text{c}=0.5R_\text{s}$, and $\lambda/(E_\text{s}R_\text{s})=0.1$.
The colour code reflects the three scaled components of the symmetric stress tensor $\sigma_{rr}^\text{(i)}/E_\text{i}$, $|\sigma_{r\theta}^\text{(i)}|/E_\text{i}$ and $\sigma_{\theta\theta}^\text{(i)}/E_\text{i}$ for the core $(\text{i}=c)$ and the shell $(\text{i}=s)$. Three different ratios of Young moduli $E_\text{c}/E_\text{s}$ each are shown for the three components.
The core and shell boundaries are indicated by black lines.
To achieve a better resolution, only the absolute value of  $\sigma_{r\theta}^\text{(i)}/E_\text{i}$ is shown. By symmetry,
this tensor component changes sign in the different quadrants of the $xz$-plane.
}
\label{fig:deformation_color}
\end{figure*}
\subsection{Deformational behaviour for different Poisson ratios of core and shell}

Figure \ref{fig:deformation_poisson} shows the deformational behaviour of the core and the shell as a function of their (in general different) Poisson ratios $\nu_\text{c}$ and $\nu_\text{s}$. For simplicity we here consider the same stiffness of the shell and the core, $E_\text{c}=E_\text{s}$. Moreover we fix the core size to $R_\text{c}=0.5R_\text{s}$ and the deformation amplitude to $\lambda /(E_\text{s}R_\text{s})=0.1$.

We distinguish between two different states of the displacement: I) the shell is more oblate than the core and II) the core is more oblate than the shell. In order to do so, we use the second coefficient of relative deformation of the shell $u^\text{s}_{r,2} / R_\text{s} $ and the core $u^\text{c}_{r,2} / R_\text{c} $. For state I) (reddish and greenish in Figure \ref{fig:deformation_poisson}) we have $u^\text{s}_{r,2} / R_\text{s} < u^\text{c}_{r,2} / R_\text{c} $ while, for state II) (blueish in Figure \ref{fig:deformation_poisson}) we have $u^\text{s}_{r,2} / R_\text{s} > u^\text{c}_{r,2} / R_\text{c} $. See also
the two schematic sketches on the top right-hand side of Figure \ref{fig:deformation_poisson}. The transition from I) to II), given by the same relative degree of oblate deformation $u^\text{s}_{r,2} / R_\text{s} = u^\text{c}_{r,2} / R_\text{c} $, is shown in Figure \ref{fig:deformation_poisson} by the black line separating the two regions. Remarkably, there is a non-monotonic behaviour of this line as a function of $\nu_c$ for an auxetic shell ($\nu_s \approx -0.6$) and a nearly incompressible core.

The different colour codes on the right hand side Figure \ref{fig:deformation_poisson} indicate the magnitude of the relative oblate deformation of the shell for state I) and of the core for state II). For nine selected points indicated in the $\nu_\text{c}\nu_\text{s}$-plane we show the corresponding elliptical shapes of the core and the shell as given by the components $u^\text{c}_{r}$ and $u^\text{s}_{r}$, respectively, describing the change in volume and relative oblate deformation.

At the origin in the state diagram, where $\nu_\text{c}=\nu_\text{s}=0$, the relative oblate deformation of the core and the shell are equal so that the black line passes the origin in Figure \ref{fig:deformation_poisson}. Strictly speaking, this point [and all others on the diagonal from (a) to (d)] describes a one-component system, because there the elastic properties of the core and the shell are identical.

Clearly, for the parameter combinations lying on the black line separating regions I) and II), the relative oblate deformations of core and shell are equal, as seen in Figure \ref{fig:deformation_poisson} (a), (c), {and (i)} $\left(u^\text{s}_{r,2} / R_\text{s} = u^\text{c}_{r,2} / R_\text{c}\right)$. In the special cases of (a) {and (i)} we recover spherical shapes of core and shell of changed volume $\left(u^\text{c}_{r,2} / R_\text{c} = u^\text{s}_{r,2} / R_\text{s}=0\right)$. In conclusion,  different Poisson ratios can largely
tune the behaviour of the core-shell structure under external loading.
{
\subsection{Internal stress field}
We now provide explicit data for the internal stress field. For quasi volume conserving conditions $\left(v_\text{c}=v_\text{s}=0.4999\right)$, a size ratio of $R_\text{c}/R_\text{s}=0.5$, and an amplitude of $\lambda / (E_\text{s}S_\text{s})=0.1$ of the force line density, loaded configurations of the core-shell system for three different ratios of Young moduli $E_\text{c}/E_\text{s}$ are shown in Figure \ref{fig:deformation_color}.

The loaded configurations are color coded for the components of the (symmetric) stress tensor, defined by $\underline{\sigma}^\text{(i)}=\sigma_{rr}^\text{(i)}\mathbf{e}_r \otimes \mathbf{e}_r+\sigma_{r\theta}^\text{(i)}\left(\mathbf{e}_\theta \otimes \mathbf{e}_r+\mathbf{e}_r \otimes \mathbf{e}_\theta\right)+\sigma_{\theta\theta}^\text{(i)} \mathbf{e}_\theta \otimes \mathbf{e}_\theta$,
for the core $(i=c)$ and the shell $(i=s)$. The components of the stress tensor are scaled by the respective $E_i$
in the core ($i=c$) and in the shell ($i=s$). Results for these components are calculated from Eqs.~(\ref{eqn:stress_strain}) and (\ref{sol1})-(\ref{sol21}), where the infinite series are truncated at $n=32$.

For all configurations, all components of the stress tensor are of {largest} magnitude around the equatorial line of loading the shell. Clearly, the system there experiences a displacement in positive radial (outward) direction. Due to the quasi-incompressibility of both shell and core, a strong degree of inverted displacement results at the poles.

For $E_\text{c} \ll E_\text{s}$, the soft core deforms more strongly than the surrounding harder shell and therefore experiences a higher amount of scaled stress.
The scaled stress of the quasi-incompressible shell
is transferred from the equator towards the inside by the bulk elasticity of the shell (see the right column in Figure \ref{fig:deformation_color}). Conversely, for $E_\text{c}\gg E_\text{s}$, there is hardly any influence of the deformation of the shell on the core for the scaled stresses (see the left column in Figure \ref{fig:deformation_color}). For comparison, the center column in Figure \ref{fig:deformation_color} shows a loaded one-component system $E_\text{c}=E_\text{s}$ and the corresponding scaled components of stress.  
}
\section{Conclusions}
We have analysed in detail the deformational response of an elastic core-shell system to a radially oriented force line density acting along the outside equatorial line. Natural extensions of our considerations include the following.

First, the axially symmetric situation that we addressed could be generalized to systems exposed to line densities that are modulated alone the circumference. Moreover, the effect of surface force densities applied in patches or distributed over the whole surface area could be analysed, instead of pure force line densities. In a further step, the imposed distortions may not only be imposed from outside, but could additionally result from internal active or actuation centers. Obvious candidates for corresponding actuatable cores are given by magnetic gels~\cite{weeber2018polymer,odenbach2016microstructure}. 
For these types of systems, magnetically induced deformations have already been analysed by linear elasticity theory in the case of one-component elastic spheres~\cite{fischer2019magnetostriction,fischer2020towards,fischer2020magnetically}. 

{The considered geometry of loading can effectively be realised in experiments on the mesoscale by exposing core-shell microgel particles to the interface between two immiscible fluids acting on the elastic system~\cite{Harrer2019,kolker2021interface}.} There, interfacial tension radially pulls on the equatorial circumference along the common contact line in a symmetric setup. Yet, our description can be applied to any system on any scale that can be characterized by continuum elasticity theory. For example, macroscopic elastic core-shell spheres could be generated as toy models {using} soft transparent elastic shells on an elastic core. The line of loading force could then simply be imposed by tying a cord around the equator of these macroscopic core-shell spheres and tightening it. In  this setup, the direction of the force is inverted as well. However, this in our evaluation simply means that all  directions of displacement are inverted. Such macroscopic approaches  may support the involvement of auxetic components~\cite{huang2016negative,ren2018auxetic,scarpa2004trends,lakes1987foam,chan1997fabrication,caddock1989microporous}. Depending on the materials at hand, this strategy may facilitate the experimental confirmation of our results, possibly by direct visual inspection.


\section*{Acknowledgements}
L.F. and A.M.M. thank the Deutsche Forschungsgemeinschaft (DFG) for support through the SPP 1681 on magnetic hybrid materials, grant no.~ME 3571/2-3, and for support through the Heisenberg Grant ME 3571/4-1 (A.M.M.). H.L. acknowledges funding from the Deutsche Forschungsgemeinschaft (DFG) under grant number LO 418/22-1.





\twocolumn[
\begin{@twocolumnfalse}
\section{Supporting Information} 

\pagebreak

In this supporting information, the Navier-Cauchy equations and stress-strain relations {Eqs.~(1) and (2) in the main text, respectively,} are presented in spherical coordinates for the problem under investigation. 
The further dependences of the coefficients $a_n^\text{c}$, $b_n^\text{c}$, $a_n^\text{s}$, $b_n^\text{s}$, $c_n^\text{s}$, $d_n^\text{s}$ on the dimensionless parameters 
$\frac{\lambda}{E_\text{s}R_\text{s}}$, $\frac{E_\text{c}}{E_\text{s}}$, $\frac{R_\text{c}}{R_\text{s}}$, $\nu_\text{c}$, $\nu_\text{s}$ {and on the index $n$} are listed in a two-step order. 
First, the dependence of the coefficients on the amplitude of the deformation $\frac{\lambda}{E_\text{s}R_\text{s}}$, the ratio of the Young moduli $\frac{E_\text{c}}{E_\text{s}}$ and the ratio of the radii $\frac{R_\text{c}}{R_\text{s}}$ is shown and in the second step the dependence on the index $n$ as well as on the Poisson ratios for core $\nu_\text{c}$ and shell $\nu_\text{s}$ is emphasised. 
At last the asymptotic behaviour for $n\rightarrow\infty$ is analysed for the Legendre polynomials and {the general rescaled solutions for the radial component of the displacement field for the core and the shell.}
\subsection{Navier-Cauchy equations and stress-strain relations in spherical coordinates}
Due to the special axial symmetry of the problem, the azimuthal component $u_\phi$ of the displacement field $\mathbf{u}\left(\mathbf{r}\right)$ is zero and any $\phi$-dependence {vanishes}. Therefore, the displacement field can be written as $\mathbf{u}\left(\mathbf{r}\right)=u_r(r,\theta)\mathbf{e}_r+u_\theta(r,\theta)\mathbf{e}_\theta${, where $\mathbf{e}_r$ and $\mathbf{e}_\theta$ denote the radial and polar unit vectors, respectively. Then} the homogeneous Navier-Cauchy equations{, Eq.~(1) in the main text,}  in spherical coordinates for the problem under investigation {become} 
\begin{align}
{
}
2(1-\nu)\left(\frac{\partial}{\partial r}\left(\frac{1}{r^2}\frac{\partial}{\partial r}\left(r^2u_r\left(r,\theta\right)\right)\right)+\frac{1}{\sin\theta}\frac{\partial}{\partial r}\left(\frac{1}{r}\frac{\partial}{\partial \theta}\left(\sin\theta u_\theta\left(r,\theta\right)\right)\right)\right)&\nonumber\\
-(1-2\nu)\left(\frac{1}{r^2\sin\theta}\frac{\partial}{\partial \theta}\left(\sin\theta\left(\frac{\partial}{\partial r}\left(ru_\theta\left(r,\theta\right)\right)-\frac{\partial}{\partial \theta}u_r\left(r,\theta\right)\right)\right)\right)&=0
\end{align}
{for the radial direction and} 
\begin{align}
2(1-\nu)\left(\frac{1}{r^3}\frac{\partial}{\partial \theta}\frac{\partial}{\partial r}\left(r^2u_r\left(r,\theta\right)\right)+\frac{1}{r^2}\frac{\partial}{\partial \theta}\left(\frac{1}{\sin\theta}\frac{\partial}{\partial \theta}\left(\sin\theta u_\theta\left(r,\theta\right)\right)\right)\right)&\nonumber\\
-(1-2\nu)\left(-\frac{1}{r}\frac{\partial}{\partial r}\left(\frac{\partial}{\partial r}\left(ru_\theta\left(r,\theta\right)\right)-\frac{\partial}{\partial\theta}u_r\left(r,\theta\right)\right)\right)&=0
\end{align}
{for the polar direction. The nontrivial components of the stress-strain relation, Eq.~(2) in the main text,} in spherical coordinates for the underlying problem {read}
\begin{align}
\sigma_{rr}(r,\theta)=&\frac{E}{1+\nu}\left(\varepsilon_{rr}(r,\theta)+\frac{\nu}{1-2\nu}\left(\varepsilon_{rr}(r,\theta)+\varepsilon_{\theta\theta}(r,\theta)\right)\right),\\
\sigma_{r\theta}(r,\theta)=&\frac{E}{1+\nu}\varepsilon_{r\theta}(r,\theta),\\
\sigma_{\theta\theta}(r,\theta)=&\frac{E}{1+\nu}\left(\varepsilon_{\theta\theta}(r,\theta)+\frac{\nu}{1-2\nu}\left(\varepsilon_{rr}(r,\theta)+\varepsilon_{\theta\theta}(r,\theta)\right)\right).
\end{align}
{Here, in} spherical coordinates, we inserted for the symmetric Cauchy stress tensor $\underline{\boldsymbol{\sigma}}\left(\mathbf{r}\right)=\sigma_{rr}(r,\theta)\mathbf{e}_r\otimes \mathbf{e}_r+\sigma_{r\theta}(r,\theta)\left(\mathbf{e}_\theta\otimes \mathbf{e}_r+\mathbf{e}_r\otimes \mathbf{e}_\theta\right)+\sigma_{\theta\theta}(r,\theta)\mathbf{e}_\theta\otimes \mathbf{e}_\theta$ and for the strain tensor $\underline{\boldsymbol{\varepsilon}}\left(\mathbf{r}\right)=\varepsilon_{rr}(r,\theta)\mathbf{e}_r\otimes \mathbf{e}_r+\varepsilon_{r\theta}(r,\theta)\left(\mathbf{e}_\theta\otimes \mathbf{e}_r+\mathbf{e}_r\otimes \mathbf{e}_\theta\right)+\varepsilon_{\theta\theta}(r,\theta)\mathbf{e}_\theta\otimes \mathbf{e}_\theta${, where $\otimes$ denotes the dyadic product.}
\end{@twocolumnfalse}
]
\twocolumn[
\begin{@twocolumnfalse}
\subsection{Dependence of the coefficients $a_n^\text{c},b_n^\text{c},a_n^\text{s},b_n^\text{s},c_n^\text{s},d_n^\text{s}$ on the amplitude of deformation $\frac{\lambda}{E_\text{s}R_\text{s}}$, the ratio of Young moduli $\frac{E_\text{c}}{E_\text{s}}$ and the ratio of radii $\frac{R_\text{c}}{R_\text{s}}$}
The coefficients $a_n^\text{c},b_n^\text{c},a_n^\text{s},b_n^\text{s},c_n^\text{s},d_n^\text{s}$ with $n \geq 0$
are listed and their dependence on the amplitude of deformation $\frac{\lambda}{E_\text{s}R_\text{s}}$, the ratio of Young moduli $\frac{E_\text{c}}{E_\text{s}}$ and the ratio of radii $\frac{R_\text{c}}{R_\text{s}}$ is highlighted. The expressions are found from the solutions of the relative deformation $\frac{\mathbf{u}^\text{c}\left(R_\text{c}\mathbf{e}_r\right)}{R_\text{c}}$ and $\frac{\mathbf{u}^\text{s}\left(R_\text{s}\mathbf{e}_r\right)}{R_\text{s}}$, respectively:
{
\begin{align*}
\frac{a^\text{c}_n}{R_\text{c}}R_\text{c}^{n+1}=&\frac{\lambda}{E_\text{s}R_\text{s}}\frac{2n+1}{2}P_n\left(0\right)\left(\frac{R_\text{c}}{R_\text{s}}\right)^{(n-2)}
\left[\left(\frac{E_\text{c}}{E_\text{s}}\right)\tilde{c}_{01,n}+\tilde{c}_{02,n}\right]\frac{1}{D},
\end{align*}
\begin{align*}
\frac{b^\text{c}_n}{R_\text{c}}R_\text{c}^{n-1}=&-\frac{\lambda}{E_\text{s}R_\text{s}}\frac{2n+1}{2}P_n\left(0\right)\left(\frac{R_\text{c}}{R_\text{s}}\right)^{(n-2)}\left[\left(\frac{E_\text{c}}{E_\text{s}}\right)\tilde{c}_{03,n}+\tilde{c}_{04,n}\right]\frac{1}{D},
\end{align*}
\begin{align*}
\frac{a^\text{s}_n}{R_\text{s}}R_\text{s}^{n+1}=&\frac{\lambda}{E_\text{s}R_\text{s}}\frac{2n+1}{2}P_n\left(0\right)
\left[\left(\frac{E_\text{c}}{E_\text{s}}\right)^2\tilde{c}_{05,n}+\left(\frac{E_\text{c}}{E_\text{s}}\right)\tilde{c}_{06,n}+\tilde{c}_{07,n}\right]\frac{1}{D},\\
\frac{b^\text{s}_n}{R_\text{s}}R_\text{s}^{n-1}=&-\frac{\lambda}{E_\text{s}R_\text{s}}\frac{2n+1}{2}P_n\left(0\right)
\left[\left(\frac{E_\text{c}}{E_\text{s}}\right)^2\tilde{c}_{08,n}+\left(\frac{E_\text{c}}{E_\text{s}}\right)\tilde{c}_{09,n}+\tilde{c}_{10,n}\right]\frac{1}{D},\\
\frac{c^\text{s}_n}{R_\text{s}}R_\text{s}^{-n}=&\frac{\lambda}{E_\text{s}R_\text{s}}\frac{2n+1}{2}P_n\left(0\right)\left(\frac{R_\text{c}}{R_\text{s}}\right)^{(2n-1)}\left[\left(\frac{E_\text{c}}{E_\text{s}}\right)^2\tilde{c}_{11,n}+\left(\frac{E_\text{c}}{E_\text{s}}\right)\tilde{c}_{12,n}+\tilde{c}_{13,n}\right]\frac{1}{D},\\
\frac{d^\text{s}_n}{R_\text{s}}R_\text{s}^{-(n+2)}=&-\frac{\lambda}{E_\text{s}R_\text{s}}\frac{2n+1}{2}P_n\left(0\right)\left(\frac{R_\text{c}}{R_\text{s}}\right)^{(2n+1)}\left[\left(\frac{E_\text{c}}{E_\text{s}}\right)^2\tilde{c}_{14,n}+\left(\frac{E_\text{c}}{E_\text{s}}\right)\tilde{c}_{15,n}+\tilde{c}_{16,n}\right]\frac{1}{D},
\end{align*}
}
where
\begin{align}
D=
\left(\frac{E_\text{c}}{E_\text{s}}\right)^2\tilde{c}_{17,n} +\frac{E_\text{c}}{E_\text{s}}\tilde{c}_{18,n}+\tilde{c}_{19,n}.
\label{eqn:d}
\end{align}
The constants $\tilde{c}_{01,n}$ to $\tilde{c}_{19,n}$ are given below with their dependence on the ratio of radii $\frac{R_\text{c}}{R_\text{s}}$:
\begin{align*}
\tilde{c}_{01,n}=&c_{01,n}+c_{02,n}\left(\frac{R_\text{c}}{R_\text{s}}\right)^{2}+c_{03,n}\left(\frac{R_\text{c}}{R_\text{s}}\right)^{(2n+1)}+c_{04,n}\left(\frac{R_\text{c}}{R_\text{s}}\right)^{(2n+3)},\\
\tilde{c}_{02,n}=&c_{05,n}+c_{06,n}\left(\frac{R_\text{c}}{R_\text{s}}\right)^{2}+c_{07,n}\left(\frac{R_\text{c}}{R_\text{s}}\right)^{(2n+1)}+c_{08,n}\left(\frac{R_\text{c}}{R_\text{s}}\right)^{(2n+3)},\\
\tilde{c}_{03,n}=&c_{09,n}+c_{10,n}\left(\frac{R_\text{c}}{R_\text{s}}\right)^{2}+c_{11,n}\left(\frac{R_\text{c}}{R_\text{c}}\right)^{(2n+1)}+c_{12,n}\left(\frac{R_\text{c}}{R_\text{s}}\right)^{(2n+3)},\\
\tilde{c}_{04,n}=&c_{13,n}+c_{14,n}\left(\frac{R_\text{c}}{R_\text{s}}\right)^{2}+c_{15,n}\left(\frac{R_\text{c}}{R_\text{s}}\right)^{(2n+1)}+c_{16,n}\left(\frac{R_\text{c}}{R_\text{s}}\right)^{(2n+3)},\\
\tilde{c}_{05,n}=&c_{17,n}+c_{18,n}\left(\frac{R_\text{c}}{R_\text{s}}\right)^{(2n-1)}+c_{19,n}\left(\frac{R_\text{c}}{R_\text{s}}\right)^{(2n+1)},\\
\tilde{c}_{06,n}=&c_{20,n}+c_{21,n}\left(\frac{R_\text{c}}{R_\text{s}}\right)^{(2n-1)}+c_{22,n}\left(\frac{R_\text{c}}{R_\text{s}}\right)^{(2n+1)},\\
\tilde{c}_{07,n}=&c_{23,n}+c_{24,n}\left(\frac{R_\text{c}}{R_\text{s}}\right)^{(2n-1)}+c_{25,n}\left(\frac{R_\text{c}}{R_\text{s}}\right)^{(2n+1)},
\end{align*}
\end{@twocolumnfalse}
]
\twocolumn[
\begin{@twocolumnfalse}
\begin{align*}
\tilde{c}_{08,n}=&c_{26,n}+c_{27,n}\left(\frac{R_\text{c}}{R_\text{s}}\right)^{(2n+1)}+c_{28,n}\left(\frac{R_\text{c}}{R_\text{s}}\right)^{(2n+3)},\\
\tilde{c}_{09,n}=&c_{29,n}+c_{30,n}\left(\frac{R_\text{c}}{R_\text{s}}\right)^{(2n+1)}+c_{31,n}\left(\frac{R_\text{c}}{R_\text{s}}\right)^{(2n+3)},\\
\tilde{c}_{10,n}=&c_{32,n}+c_{33,n}\left(\frac{R_\text{c}}{R_\text{s}}\right)^{(2n+1)}+c_{34,n}\left(\frac{R_\text{c}}{R_\text{s}}\right)^{(2n+3)},\\
\tilde{c}_{11,n}=&c_{35,n}+c_{36,n}\left(\frac{R_\text{c}}{R_\text{s}}\right)^{2}+c_{37,n}\left(\frac{R_\text{c}}{R_\text{s}}\right)^{(2n+3)},\\
\tilde{c}_{12,n}=&c_{38,n}+c_{39,n}\left(\frac{R_\text{c}}{R_\text{s}}\right)^{2}+c_{40,n}\left(\frac{R_\text{c}}{R_\text{s}}\right)^{(2n+3)},\\
\tilde{c}_{13,n}=&c_{41,n}+c_{42,n}\left(\frac{R_\text{c}}{R_\text{s}}\right)^{2}+c_{43,n}\left(\frac{R_\text{c}}{R_\text{s}}\right)^{(2n+3)},\\
\tilde{c}_{14,n}=&c_{44,n}+c_{45,n}\left(\frac{R_\text{c}}{R_\text{s}}\right)^{2}+c_{46,n}\left(\frac{R_\text{c}}{R_\text{s}}\right)^{(2n+1)},\\
\tilde{c}_{15,n}=&c_{47,n}+c_{48,n}\left(\frac{R_\text{c}}{R_\text{s}}\right)^{2}+c_{49,n}\left(\frac{R_\text{c}}{R_\text{s}}\right)^{(2n+1)},\\
\tilde{c}_{16,n}=&c_{50,n}+c_{51,n}\left(\frac{R_\text{c}}{R_\text{s}}\right)^{2}+c_{52,n}\left(\frac{R_\text{c}}{R_\text{s}}\right)^{(2n+1)},\\
\tilde{c}_{17,n}=&c_{53,n}+c_{54,n}\left(\frac{R_\text{c}}{R_\text{s}}\right)^{(2n-1)}+c_{55,n}\left(\frac{R_\text{c}}{R_\text{s}}\right)^{(2n+1)}
+c_{56,n}\left(\frac{R_\text{c}}{R_\text{s}}\right)^{(2n+3)}+c_{57,n}\left(\frac{R_\text{c}}{R_\text{s}}\right)^{(4n+2)},\\
\tilde{c}_{18,n}=&c_{58,n}+c_{59,n}\left(\frac{R_\text{c}}{R_\text{s}}\right)^{(2n-1)}+c_{60,n}\left(\frac{R_\text{c}}{R_\text{s}}\right)^{(2n+1)}+c_{61,n}\left(\frac{R_\text{c}}{R_\text{s}}\right)^{(2n+3)}+c_{62,n}\left(\frac{R_\text{c}}{R_\text{s}}\right)^{(4n+2)},\\
\tilde{c}_{19,n}=&c_{63,n}+c_{64,n}\left(\frac{R_\text{c}}{R_\text{s}}\right)^{(2n-1)}+c_{65,n}\left(\frac{R_\text{c}}{R_\text{s}}\right)^{(2n+1)}+c_{66,n}\left(\frac{R_\text{c}}{R_\text{s}}\right)^{(2n+3)}+c_{67,n}\left(\frac{R_\text{c}}{R_\text{s}}\right)^{(4n+2)}.\\
\end{align*}
\end{@twocolumnfalse}
]
\twocolumn[
\begin{@twocolumnfalse}
\subsection{Dependence of the constants $c_{01,n}$ to $c_{67,n}$ on the index $n$, the Poisson ratio of the core $\nu_\text{c}$ and of the shell $\nu_\text{s}$}
The constants $c_{01,n}$ to $c_{67,n}$ only depend on the index $n$, the Poisson ratio of the core $\nu_\text{c}$ and of the shell $\nu_\text{s}$. They are listed below:
{
\begin{align*}
c_{01,n}&=0,\\
c_{02,n}&=-\frac{4 (-1 + n)^2 (3 + 8 n + 4 n^2) (-1 + \nu_\text{s}) (-2 - 3 n + 2 \nu_\text{s} + 
    4 n \nu_\text{s})}{(1 + \nu_\text{c}) (1 + \nu_\text{s})^2},\\
c_{03,n}&=-\frac{2 (1 + 2 n)^2 (-3 + n + 2 n^2) (-1 + \nu_\text{s}) (-2 + n^2 + 2 \nu_\text{s})}{(1 + 
    \nu_\text{c}) (1 + \nu_\text{s})^2},\\
c_{04,n}&=\frac{2 n (2 + n) (3 - n - 14 n^2 + 4 n^3 + 8 n^4) (-1 + \nu_\text{s})}{(1 + 
   \nu_\text{c}) (1 + \nu_\text{s})^2},\\
c_{05,n}&=0,\\
c_{06,n}&=\frac{4 (-1 + n) (3 + 8 n + 4 n^2) (-1 + \nu_\text{s}) (1 + n + n^2 - \nu_\text{s} - 
   2 n \nu_\text{s})}{(1 + \nu_\text{s})^3},\\
c_{07,n}&=\frac{2 (1 + 2 n)^2 (-3 + n + 2 n^2) (-1 + \nu_\text{s}) (-2 + n^2 + 2 \nu_\text{s})}{(1 + \nu_\text{s})^3},\\
c_{08,n}&=-\frac{2 n (2 + n) (3 - n - 14 n^2 + 4 n^3 + 8 n^4) (-1 + \nu_\text{s})}{(1 + \nu_\text{s})^3},\\
c_{09,n}&=\frac{4 (-1 + 4 n^2) (1 + n + n^2 + \nu_\text{c} + 2 n \nu_\text{c}) (-1 + \nu_\text{s}) (-1 + 2 n + 
   n^2 + 2 \nu_\text{s})}{(1 + \nu_\text{c}) (1 + \nu_\text{s})^2},\\
c_{10,n}&=-\frac{4 (-1 + n) (3 + 11 n + 12 n^2 + 4 n^3) (-1 + \nu_\text{s}) (5 - \nu_\text{c} - 6 \nu_\text{s} + 
    2 n (-1 + \nu_\text{c} + \nu_\text{s}) + n^2 (-2 + 4 \nu_\text{s}))}{(1 + \nu_\text{c}) (1 + \nu_\text{s})^2},\\
c_{11,n}&=-\frac{2 (3 + 2 n)^2 (-1 - 2 n + n^2 + 2 n^3) (-1 + \nu_\text{s}) (-2 + n^2 + 
    2 \nu_\text{s})}{(1 + \nu_\text{c}) (1 + \nu_\text{s})^2},\\
c_{12,n}&=2 (2 + n) (-1 + 4 n^2) (-1 + \nu_\text{s}) \left[\frac{5 n^3 + 2 n^4 + n^2 (6 - 8 \nu_\text{s})}{(1 + 
   \nu_\text{c}) (1 + \nu_\text{s})^2}\right.\\
    &+\left.\frac{ 
   -4 (1 + \nu_\text{c}) (-1 + 2 \nu_\text{s}) - n (1 + 8 \nu_\text{s} + 4 \nu_\text{c} (-3 + 4 \nu_\text{s}))}{(1 + 
   \nu_\text{c}) (1 + \nu_\text{s})^2}\right],\\
c_{13,n}&=-\frac{4 (-2 - n + 8 n^2 + 4 n^3) (-1 + 2 \nu_\text{c} + n (-3 + 4 \nu_\text{c})) (-1 + 
    \nu_\text{s}) (-1 + 2 n + n^2 + 2 \nu_\text{s})}{(1 + \nu_\text{s})^3},\\
c_{14,n}&=\frac{4 (-1 + n) (3 + 11 n + 12 n^2 + 4 n^3) (-1 + \nu_\text{s}) (5 - 4 \nu_\text{c} + 
   n^2 (-2 + 4 \nu_\text{c}) - 3 \nu_\text{s} + n (6 \nu_\text{c} - 2 (1 + \nu_\text{s})))}{(1 + \nu_\text{s})^3},\\
c_{15,n}&=\frac{2 (3 + 2 n)^2 (-1 - 2 n + n^2 + 2 n^3) (-1 + \nu_\text{s}) (-2 + n^2 + 
   2 \nu_\text{s})}{(1 + \nu_\text{s})^3},\\
c_{16,n}&=-2 (2 + n) (-1 + 4 n^2) (-1 + \nu_\text{s}) \left[\frac{5 n^3 + 2 n^4 + n^2 (6 - 8 \nu_\text{c})}{(1 + 
   \nu_\text{s})^3}\right.\\
&+\left.\frac{ - 
    4 (-1 + 2 \nu_\text{c}) (1 + \nu_\text{s}) - n (1 - 12 \nu_\text{s} + 8 \nu_\text{c} (1 + 2 \nu_\text{s}))}{(1 + 
   \nu_\text{s})^3}\right],
      \end{align*}
   }
\end{@twocolumnfalse}
]
\twocolumn[
\begin{@twocolumnfalse}
{
\begin{align*}
c_{17,n}&=\frac{4 (-1 + n)^2 (1 + n + n^2 + \nu_\text{c} + 2 n \nu_\text{c}) (-2 - 3 n + 2 \nu_\text{s} + 
   4 n \nu_\text{s})}{(1 + \nu_\text{c})^2 (1 + \nu_\text{s})},\\
c_{18,n}&=\frac{2 (-1 + n) (1 + 2 n) (1 + n + n^2 + \nu_\text{c} + 2 n \nu_\text{c}) (-2 + n^2 + 
   2 \nu_\text{s})}{(1 + \nu_\text{c})^2 (1 + \nu_\text{s})},\\
c_{19,n}&=-\frac{2 (-1 + n) n (2 + n) (-1 + 2 n) (1 + n + n^2 + \nu_\text{c} + 2 n \nu_\text{c})}{(1 + 
    \nu_\text{c})^2 (1 + \nu_\text{s})},\\
c_{20,n}&=-4 (-1 + n)\left[\frac{ -3 (-1 + 3 \nu_\text{c}) (-1 + \nu_\text{s}) + 
    n^2 (-4 + \nu_\text{c} (9 - 16 \nu_\text{s}) + 9 \nu_\text{s})}{(1 + \nu_\text{c}) (1 + \nu_\text{s})^2} \right.\\ 
    &\left.+\frac{n (-14 + \nu_\text{c} (27 - 32 \nu_\text{s}) + 15 \nu_\text{s}) + 
    4 n^3 (5 - 6 \nu_\text{s} + \nu_\text{c} (-6 + 8 \nu_\text{s})) + 
    2 n^4 (5 - 6 \nu_\text{s} + \nu_\text{c} (-6 + 8 \nu_\text{s}))}{(1 + \nu_\text{c}) (1 + \nu_\text{s})^2}\right],\\
    c_{21,n}&=-\frac{2 (-1 - n + 2 n^2) (-1 + 5 \nu_\text{c} + 6 n (-1 + 2 \nu_\text{c}) + 
    n^2 (-2 + 4 \nu_\text{c})) (-2 + n^2 + 2 \nu_\text{s})}{(1 + \nu_\text{c}) (1 + \nu_\text{s})^2},\\
c_{22,n}&=\frac{2 n (2 + n) (1 - 3 n + 2 n^2) (-1 + 5 \nu_\text{c} + 6 n (-1 + 2 \nu_\text{c}) + 
   n^2 (-2 + 4 \nu_\text{c}))}{(1 + \nu_\text{c}) (1 + \nu_\text{s})^2},\\
c_{23,n}&=\frac{4 (-1 + n) (2 + n) (-1 + 2 \nu_\text{c} + n (-3 + 4 \nu_\text{c})) (1 + n + n^2 - \nu_\text{s} - 
   2 n \nu_\text{s})}{(1 + \nu_\text{s})^3},\\
c_{24,n}&=\frac{2 (-2 - 3 n + 3 n^2 + 2 n^3) (-1 + 2 \nu_\text{c} + n (-3 + 4 \nu_\text{c})) (-2 + n^2 + 
   2 \nu_\text{s})}{(1 + \nu_\text{s})^3},\\
c_{25,n}&=-\frac{2 n (2 + n)^2 (1 - 3 n + 2 n^2) (-1 + 2 \nu_\text{c} + n (-3 + 4 \nu_\text{c}))}{(1 + 
   \nu_\text{s})^3},\\
c_{26,n}&=\frac{4 (-1 + n) (1 + n + n^2 + \nu_\text{c} + 2 n \nu_\text{c}) (-2 - 3 n + 2 \nu_\text{s} + 
   4 n \nu_\text{s}) (-1+2n+n^2+2\nu_\text{s})}{(1 + \nu_\text{s})(1 + \nu_\text{c})^2},\\
c_{27,n}&=\frac{2 (-1 + n) (3 + 5 n + 2 n^2) (1 + n + n^2 + \nu_\text{c} + 2 n \nu_\text{c}) (-2 + n^2 + 
   2 \nu_\text{s})}{(1 + \nu_\text{c})^2 (1 + \nu_\text{s})}, \\
c_{28,n}&=-\frac{2 (-1 + n) (2 + n) (1 + 2 n) (1 + n + n^2 + \nu_\text{c} + 2 n \nu_\text{c}) (8 + n + 
    n^2 - 24 \nu_\text{s} + 16 \nu_\text{s}^2)}{(1 + \nu_\text{c})^2 (1 + \nu_\text{s})},\\
c_{29,n}&=-4 (-1 + 2 n + n^2 + 2 \nu_\text{s})\left[ \frac{-3 (-1 + 3 \nu_\text{c}) (-1 + \nu_\text{s}) + 
    n^2 (-4 + \nu_\text{c} (9 - 16 \nu_\text{s}) + 9 \nu_\text{s})}{(1 + \nu_\text{c}) (1 + \nu_\text{s})^2}  \right.\\
    &\left.+\frac{n (-14 + \nu_\text{c} (27 - 32 \nu_\text{s}) + 15 \nu_\text{s}) + 
    4 n^3 (5 - 6 \nu_\text{s} + \nu_\text{c} (-6 + 8 \nu_\text{s})) + 
    2 n^4 (5 - 6 \nu_\text{s} + \nu_\text{c} (-6 + 8 \nu_\text{s}))}{(1 + \nu_\text{c}) (1 + \nu_\text{s})^2}\right],\\ 
&c_{30,n}=-\frac{2 (-3 - 2 n + 3 n^2 + 2 n^3) (-1 + 5 \nu_\text{c} + 6 n (-1 + 2 \nu_\text{c}) + 
    n^2 (-2 + 4 \nu_\text{c})) (-2 + n^2 + 2 \nu_\text{s})}{(1 + \nu_\text{c}) (1 + \nu_\text{s})^2},\\
&c_{31,n}=2 (2 + n) (1 + 2 n) \left[\frac{(6 n^4 (-1 + 2 \nu_\text{c}) + 
   n^5 (-2 + 4 \nu_\text{c}) - 12 (-1 + \nu_\text{s}) (-1 + \nu_\text{c} + 2 \nu_\text{c} \nu_\text{s}) - 
   n (11 + \nu_\text{c} - 4 \nu_\text{s} - 28 \nu_\text{c} \nu_\text{s} - 8 \nu_\text{s}^2 + 32 \nu_\text{c} \nu_\text{s}^2)}{(1 + \nu_\text{c}) (1 + \nu_\text{s})^2}\right.\\
   &\left. +\frac{    n^3 (9 - 8 \nu_\text{s} + \nu_\text{c} (-15 + 16 \nu_\text{s})) + 
   2 n^2 (3 + 16 \nu_\text{s} - 16 \nu_\text{s}^2 + 2 \nu_\text{c} (-7 - 4 \nu_\text{s} + 8 \nu_\text{s}^2)))}{(1 + \nu_\text{c}) (1 + \nu_\text{s})^2}\right],\\
&c_{32,n}=\frac{4 (2 + n) (-1 + 2 \nu_\text{c} + n (-3 + 4 \nu_\text{c})) (-1 + 2 n + n^2 + 2 \nu_\text{s}) (1 + 
   n + n^2 - \nu_\text{s} - 2 n \nu_\text{s})}{(1 + \nu_\text{s})^3},
         \end{align*}
   }
\end{@twocolumnfalse}
]
\twocolumn[
\begin{@twocolumnfalse}
{
\begin{align*}
&c_{33,n}=\frac{2 (2 + n) (-3 - 2 n + 3 n^2 + 2 n^3) (-1 + 2 \nu_\text{c} + 
   n (-3 + 4 \nu_\text{c})) (-2 + n^2 + 2 \nu_\text{s})}{(1 + \nu_\text{s})^3},\\
&c_{34,n}=-\frac{2 (2 + n) (1 + 2 n) (-1 + 2 \nu_\text{c} + n (-3 + 4 \nu_\text{c})) (4 - 2 n - n^2 + 
    2 n^3 + n^4 - 4 \nu_\text{s}^2)}{(1 + \nu_\text{s})^3},\\
&c_{35,n}=-\frac{2 (-1 + n) (1 + 2 n) (1 + n + n^2 + \nu_\text{c} + 
    2 n \nu_\text{c})  (-1+2n+n^2+2\nu_\text{s})}{(1 + \nu_\text{s})(1 + \nu_\text{c})^2},\\
&c_{36,n}=\frac{2 (-1 + n)^2 (3 + 5 n + 2 n^2) (1 + n + n^2 + \nu_\text{c} + 2 n \nu_\text{c})}{(1 + 
   \nu_\text{c})^2 (1 + \nu_\text{s})},\\
&c_{37,n}=\frac{4 (-1 + n) (2 + n) (1 + n + n^2 + \nu_\text{c} + 2 n \nu_\text{c}) (-1 + 2 \nu_\text{s} + 
   n (-3 + 4 \nu_\text{s}))}{(1 + \nu_\text{c})^2 (1 + \nu_\text{s})},\\
&c_{38,n}=\frac{2 (-1 - n + 2 n^2) (-1 + 5 \nu_\text{c} + 6 n (-1 + 2 \nu_\text{c}) + 
   n^2 (-2 + 4 \nu_\text{c})) (-1 + 2 n + n^2 + 2 \nu_\text{s})}{(1 + \nu_\text{c}) (1 + \nu_\text{s})^2},\\
&c_{39,n}=-\frac{2 (-1 + n)^2 (3 + 5 n + 2 n^2) (-1 + 5 \nu_\text{c} + 6 n (-1 + 2 \nu_\text{c}) + 
    n^2 (-2 + 4 \nu_\text{c}))}{(1 + \nu_\text{c}) (1 + \nu_\text{s})^2},\\
&c_{40,n}=-\frac{4 (-1 + n) (2 + n) (-2 + \nu_\text{c} + \nu_\text{s} + 4 \nu_\text{c} \nu_\text{s} + 
    n^3 (-6 + 4 \nu_\text{c} + 4 \nu_\text{s}) + 8 n^2 (-1 + 2 \nu_\text{c} \nu_\text{s}) + 
    n (-8 + \nu_\text{c} + \nu_\text{s} + 16 \nu_\text{c} \nu_\text{s}))}{(1 + \nu_\text{c}) (1 + \nu_\text{s})^2},\\
&c_{41,n}=-\frac{2 (-2 - 3 n + 3 n^2 + 2 n^3) (-1 + 2 \nu_\text{c} + n (-3 + 4 \nu_\text{c})) (-1 + 
    2 n + n^2 + 2 \nu_\text{s})}{(1 + \nu_\text{s})^3},\\
&c_{42,n}=\frac{2 (-1 + n)^2 (2 + n) (3 + 5 n + 2 n^2) (-1 + 2 \nu_\text{c} + 
   n (-3 + 4 \nu_\text{c}))}{(1 + \nu_\text{s})^3},\\
&c_{43,n}=\frac{4 (-1 + n) (2 + n) (-1 + 2 \nu_\text{c} + n (-3 + 4 \nu_\text{c})) (1 + n + n^2 + \nu_\text{s} + 
   2 n \nu_\text{s})}{(1 + \nu_\text{s})^3},\\
&c_{44,n}=\frac{2 (-1 + n) n (-1 + 2 n) (1 + n + n^2 + \nu_\text{c} + 
   2 n \nu_\text{c})  (-1+2n+n^2+2\nu_\text{s})}{(1 + \nu_\text{s})(1 + \nu_\text{c})^2},\\
&c_{45,n}=-\frac{2 (-1 + n)^2 (1 + 2 n) (1 + n + n^2 + \nu_\text{c} + 2 n \nu_\text{c}) (8 + n + n^2 - 
    24 \nu_\text{s} + 16 \nu_\text{s}^2)}{(1 + \nu_\text{c})^2 (1 + \nu_\text{s})},\\
&c_{46,n}=-\frac{4 (-1 + n) (1 + n + n^2 + \nu_\text{c} + 2 n \nu_\text{c}) (-2 + n^2 + 2 \nu_\text{s}) (-1 + 
    2 \nu_\text{s} + n (-3 + 4 \nu_\text{s}))}{(1 + \nu_\text{c})^2 (1 + \nu_\text{s})},\\
&c_{47,n}=-\frac{2 n (1 - 3 n + 2 n^2) (-1 + 5 \nu_\text{c} + 6 n (-1 + 2 \nu_\text{c}) + 
    n^2 (-2 + 4 \nu_\text{c})) (-1 + 2 n + n^2 + 2 \nu_\text{s})}{(1 + \nu_\text{c}) (1 + \nu_\text{s})^2},\\
&c_{48,n}=2 (-1 + n) (1 + 2 n) \left[\frac{(6 n^4 (-1 + 2 \nu_\text{c}) + 
   n^5 (-2 + 4 \nu_\text{c}) - 12 (-1 + \nu_\text{s}) (-1 + \nu_\text{c} + 2 \nu_\text{c} \nu_\text{s}) - 
   n (11 + \nu_\text{c} - 4 \nu_\text{s} - 28 \nu_\text{c} \nu_\text{s} - 8 \nu_\text{s}^2 + 32 \nu_\text{c} \nu_\text{s}^2)}{(1 + \nu_\text{c}) (1 + \nu_\text{s})^2}\right. \\
   &\left.+   \frac{n^3 (9 - 8 \nu_\text{s} + \nu_\text{c} (-15 + 16 \nu_\text{s})) + 
   2 n^2 (3 + 16 \nu_\text{s} - 16 \nu_\text{s}^2 + 2 \nu_\text{c} (-7 - 4 \nu_\text{s} + 8 \nu_\text{s}^2))}{(1 + \nu_\text{c}) (1 + \nu_\text{s})^2}\right],\\   
&c_{49,n}=\frac{4 (-1 + n) (-2 + n^2 + 2 \nu_\text{s}) (-2 + \nu_\text{c} + \nu_\text{s} + 4 \nu_\text{c} \nu_\text{s} + 
   n^3 (-6 + 4 \nu_\text{c} + 4 \nu_\text{s}) + 8 n^2 (-1 + 2 \nu_\text{c} \nu_\text{s}) + 
   n (-8 + \nu_\text{c} + \nu_\text{s} + 16 \nu_\text{c} \nu_\text{s}))}{(1 + \nu_\text{c}) (1 + \nu_\text{s})^2},\\
&c_{50,n}=\frac{2 n (2 - 5 n + n^2 + 2 n^3) (-1 + 2 \nu_\text{c} + n (-3 + 4 \nu_\text{c})) (-1 + 2 n + 
   n^2 + 2 \nu_\text{s})}{(1 + \nu_\text{s})^3},
         \end{align*}
   }
\end{@twocolumnfalse}
]
\twocolumn[
\begin{@twocolumnfalse}
{
\begin{align*}
&c_{51,n}=-\frac{2 (-1 + n) (1 + 2 n) (-1 + 2 \nu_\text{c} + n (-3 + 4 \nu_\text{c})) (4 - 2 n - n^2 + 
    2 n^3 + n^4 - 4 \nu_\text{s}^2)}{(1 + \nu_\text{s})^3},\\
&c_{52,n}=-\frac{4 (-1 + n) (-1 + 2 \nu_\text{c} + n (-3 + 4 \nu_\text{c})) (-2 + n^2 + 2 \nu_\text{s}) (1 + n + 
    n^2 + \nu_\text{s} + 2 n \nu_\text{s})}{(1 + \nu_\text{s})^3},\\
&c_{53,n}=-\frac{8 (-1 + n)^2 (1 + n + n^2 + \nu_\text{c} + 2 n \nu_\text{c}) (1 + n + n^2 + \nu_\text{s} + 
    2 n \nu_\text{s}) (-2 - 3 n + 2 \nu_\text{s} + 4 n \nu_\text{s})}{(1 + \nu_\text{c})^2 (1 + \nu_\text{s})^2},\\
&c_{54,n}=\frac{2 (-1 + n) (1 + 2 n)^2 (1 + n + n^2 + \nu_\text{c} + 2 n \nu_\text{c}) (4 - 2 n - n^2 + 
   2 n^3 + n^4 - 4 \nu_\text{s}^2)}{(1 + \nu_\text{c})^2 (1 + \nu_\text{s})^2},\\
&c_{55,n}=-\frac{4 (-1 + n)^2 n (-6 - n + 17 n^2 + 16 n^3 + 4 n^4) (1 + n + n^2 + 
    \nu_\text{c} + 2 n \nu_\text{c})}{(1 + \nu_\text{c})^2 (1 + \nu_\text{s})^2},\\
&c_{56,n}=\frac{2 (-1 + n)^2 (2 + n) (1 + 2 n)^2 (1 + n + n^2 + \nu_\text{c} + 2 n \nu_\text{c}) (8 + n +
    n^2 - 24 \nu_\text{s} + 16 \nu_\text{s}^2)}{(1 + \nu_\text{c})^2 (1 + \nu_\text{s})^2},\\
&c_{57,n}=-\frac{8 (-1 + n) (2 + n) (1 + n + n^2 + \nu_\text{c} + 2 n \nu_\text{c}) (1 + n + n^2 - \nu_\text{s} - 
    2 n \nu_\text{s}) (-1 + 2 \nu_\text{s} + n (-3 + 4 \nu_\text{s}))}{(1 + \nu_\text{c})^2 (1 + \nu_\text{s})^2},\\
&c_{58,n}=8 (-1 + n) (1 + n + n^2 + \nu_\text{s} + 2 n \nu_\text{s}) \left[\frac{-3 (-1 + 3 \nu_\text{c}) (-1 + \nu_\text{s}) + 
   n^2 (-4 + \nu_\text{c} (9 - 16 \nu_\text{s}) + 9 \nu_\text{s})}{(1 + \nu_\text{c}) (1 + \nu_\text{s})^3} \right.\\
   &\left.+ \frac{
   n (-14 + \nu_\text{c} (27 - 32 \nu_\text{s}) + 15 \nu_\text{s}) + 
   4 n^3 (5 - 6 \nu_\text{s} + \nu_\text{c} (-6 + 8 \nu_\text{s})) + 
   2 n^4 (5 - 6 \nu_\text{s} + \nu_\text{c} (-6 + 8 \nu_\text{s}))}{(1 + \nu_\text{c}) (1 + \nu_\text{s})^3}\right],\\
&c_{59,n}=-\frac{2 (-1 + n) (1 + 2 n)^2 (-1 + 5 \nu_\text{c} + 6 n (-1 + 2 \nu_\text{c}) + 
    n^2 (-2 + 4 \nu_\text{c})) (4 - 2 n - n^2 + 2 n^3 + n^4 - 4 \nu_\text{s}^2)}{(1 + 
    \nu_\text{c}) (1 + \nu_\text{s})^3},\\
&c_{60,n}=\frac{4 (-1 + n)^2 n (-6 - n + 17 n^2 + 16 n^3 + 4 n^4) (-1 + 5 \nu_\text{c} + 
   6 n (-1 + 2 \nu_\text{c}) + n^2 (-2 + 4 \nu_\text{c}))}{(1 + \nu_\text{c}) (1 + \nu_\text{s})^3},\\
&c_{61,n}=-2 (1 + 2 n)^2 (-2 + n + n^2)\left[\frac{ 6 n^4 (-1 + 2 \nu_\text{c}) + n^5 (-2 + 4 \nu_\text{c}) - 
    12 (-1 + \nu_\text{s}) (-1 + \nu_\text{c} + 2 \nu_\text{c} \nu_\text{s}) }{(1 + \nu_\text{c}) (1 + \nu_\text{s})^3}\right.\\
    &\left.+ \frac{-
    n (11 + \nu_\text{c} - 4 \nu_\text{s} - 28 \nu_\text{c} \nu_\text{s} - 8 \nu_\text{s}^2 + 32 \nu_\text{c} \nu_\text{s}^2) + 
    n^3 (9 - 8 \nu_\text{s} + \nu_\text{c} (-15 + 16 \nu_\text{s})) + 
    2 n^2 (3 + 16 \nu_\text{s} - 16 \nu_\text{s}^2 + 2 \nu_\text{c} (-7 - 4 \nu_\text{s} + 8 \nu_\text{s}^2))}{(1 + \nu_\text{c}) (1 + \nu_\text{s})^3}\right],\\
&c_{62,n}=8 (-1 + n) (2 + n)(1 + n + n^2 - \nu_\text{s} - 2 n \nu_\text{s})\left[ \frac{ -2 + \nu_\text{c} + \nu_\text{s} + 
   4 \nu_\text{c} \nu_\text{s} + n^3 (-6 + 4 \nu_\text{c} + 4 \nu_\text{s}) + 8 n^2 (-1 + 2 \nu_\text{c} \nu_\text{s})}{(1 + \nu_\text{c}) (1 + \nu_\text{s})^3} \right.\\
   &\left.+ 
  \frac{ n (-8 + \nu_\text{c} + \nu_\text{s} + 16 \nu_\text{c} \nu_\text{s})}{(1 + \nu_\text{c}) (1 + \nu_\text{s})^3}\right],\\
&c_{63,n}=-\frac{8 (-1 + n) (2 + n) (-1 + 2 \nu_\text{c} + n (-3 + 4 \nu_\text{c})) (1 + n + n^2 - \nu_\text{s} - 
    2 n \nu_\text{s}) (1 + n + n^2 + \nu_\text{s} + 2 n \nu_\text{s})}{(1 + \nu_\text{s})^4},\\
&c_{64,n}=\frac{2 (1 + 2 n)^2 (-2 + n + n^2) (-1 + 2 \nu_\text{c} + n (-3 + 4 \nu_\text{c})) (4 - 2 n - 
   n^2 + 2 n^3 + n^4 - 4 \nu_\text{s}^2)}{(1 + \nu_\text{s})^4},\\
&c_{65,n}=-\frac{4 n (-2 + n + n^2)^2 (-3 + n + 8 n^2 + 4 n^3) (-1 + 2 \nu_\text{c} + 
    n (-3 + 4 \nu_\text{c}))}{(1 + \nu_\text{s})^4},\\
&c_{66,n}=\frac{2 (-1 + n) (2 + n) (1 + 2 n)^2 (-1 + 2 \nu_\text{c} + n (-3 + 4 \nu_\text{c})) (4 - 2 n -
    n^2 + 2 n^3 + n^4 - 4 \nu_\text{s}^2)}{(1 + \nu_\text{s})^4},
             \end{align*}
   }
\end{@twocolumnfalse}
]
\twocolumn[
\begin{@twocolumnfalse}
{
\begin{align*}
&c_{67,n}=-\frac{8 (-1 + n) (2 + n) (-1 + 2 \nu_\text{c} + n (-3 + 4 \nu_\text{c})) (1 + n + n^2 - \nu_\text{s} - 
    2 n \nu_\text{s}) (1 + n + n^2 + \nu_\text{s} + 2 n \nu_\text{s})}{(1 + \nu_\text{s})^4}.\\   
\end{align*}
}
\subsection{Asymptotic behaviour of the Legendre polynomials $P_n$ and the {general rescaled solutions for the radial component of the displacement field for the core $u_{r}^\text{(c)}/R_c$ and for the shell $u_{r}^\text{(s)}/R_s$}}
For the Legendre polynomials $P_n\left(\cos\theta\right)$ with $\theta=\pi/2$, the dependence on the index $n$ is as follows~\cite{arfken2005mathematical}
\begin{align}
P_n\left(0\right)=
\begin{cases}
\frac{(-1)^m}{2^{2m}}\frac{(2m)!}{(m!)^2} &\text{for}\,\,\,\, n=2m, \\
0 &\text{for}\,\,\,\, n=2m+1.\\
\end{cases}
\end{align} 
Let $a_m=\frac{1}{2^{2m}}\frac{(2m)!}{(m!)^2}$. To calculate the asymptotic behaviour of this coefficient for $m\rightarrow \infty$, Stirling's formula can be used:
\begin{align}
N!=\sqrt{2\pi N}\left(\frac{N}{e}\right)^N\left(1+\mathcal{O}\left(\frac{1}{N}\right)\right),
\end{align}
{where $e$ denotes Euler's number.} Applying this formula to $a_m$ leads (for large $m$) to:
\begin{align}
a_m &\approx \frac{1}{2^{2m}}\frac{\sqrt{2\pi 2m}}{2\pi m}\left(\frac{2m}{e}\right)^{2m}\left(\frac{e}{m}\right)^{2m}\nonumber\\
&=\frac{2^{2m}}{2^{2m}}\frac{\sqrt{2\pi 2m}}{2\pi m}\nonumber\\
&=\frac{1}{\sqrt{\pi m}}.
\end{align}
Thus, the following asymptotic behaviour for $P_n(0)$ results:
\begin{align}
P_n(0)\approx
\begin{cases}
\sqrt{\frac{2}{\pi n}} & \text{for}\,\,\,\, n \,\text{even},\\
0 & \text{for}\,\,\,\, n \,\text{odd}.\\
\end{cases}
\end{align}
Furthermore, the dependence on the index $n$ for the angles $\theta=0,\pi$ gives~\cite{arfken2005mathematical}
\begin{align}
&P_n\left(1\right)=1,\\
&P_n\left(-1\right)=
\begin{cases}
1 & \text{for}\,\,\,\, n \,\text{even},\\
-1 & \text{for}\,\,\,\, n \,\text{odd}.\\
\end{cases}
\end{align} 
The case {of $n$ being} odd is, due to the assumed mirror symmetry, irrelevant for the investigated problem, therefore $P_n(\cos 0)=P_n(\cos \pi)=1$ holds true.

The general {rescaled solution} for the radial component of the displacement field for the core $u_{r}^\text{(c)}/R_c$ is obtained at the core radius $R_c$ as follows
\begin{align}
\frac{u_{r}^\text{(c)}(R_c\mathbf{e}_r)}{R_c}=\sum_{n=0}^\infty & G^\text{(c)}_{r,n}\left(\frac{\lambda}{E_sR_s},\frac{E_c}{E_s}, \frac{R_c}{R_s},\nu_c,\nu_s\right)\frac{2n+1}{2}P_n\left(0\right)P_n\left(\cos\theta\right)\label{eqn:gc2}
\end{align} 
where {$G^\text{(c)}_{r,n}\left(\lambda/(E_sR_s),E_c/E_s, R_c/R_s,\nu_c,\nu_s\right)$ is the corresponding kernel function of the core} and the {remaining} factors in the sum result from the expansion of the {Dirac delta function} in Legendre polynomials. {In terms of the coefficients $a_n^\text{c}$ and $b_n^\text{c}$, (\ref{eqn:gc2}) can also be written as}
\end{@twocolumnfalse}
]
\twocolumn[
\begin{@twocolumnfalse}
\begin{align}
\frac{u_{r}^\text{(c)}\left(R_c\mathbf{e}_r\right)}{R_c}=\sum_{n=0}^\infty & \left(\frac{a_n^\text{c}}{R_c}R_c^{n+1}(n+1)(-2+n+4\nu_c)+\frac{b_n^\text{c}}{R_c}R_c^{n-1}n\right)P_n\left(\cos \theta\right)\nonumber\\
=\sum_{n=0}^\infty & \frac{2n+1}{2}P_n(0)P_n\left(\cos \theta\right)\left(\frac{R_c}{R_s}\right)^{(n-2)}\frac{\lambda}{E_sR_s}\nonumber\\
&\times\frac{1}{D}\left(\underbrace{\left[\left(\frac{E_c}{E_s}\right)\tilde{c}_{01,n}+\tilde{c}_{02,n}\right](n+1)(-2+n+4\nu_c)}_{I}\underbrace{-\left[\left(\frac{E_c}{E_s}\right)\tilde{c}_{03,n}+\tilde{c}_{04,n}\right]n}_{II}\right)\label{equ_gc}.
\end{align}
Comparing the solution for $u_{r}^\text{(c)}/R_c$ here with that in (\ref{eqn:gc2}), it can be concluded that the {kernel function} of the core $G_{r, n}^\text{(c)}$ is the product of the factors $(R_c/R_s)^{(n-2)}$, $\lambda/(E_sR_s)$, $1/D$ {(see Eq.~(\ref{eqn:d}))} and the sum of \ref{equ_gc}I + \ref{equ_gc}II. By multiplying the sum \ref{equ_gc}I + \ref{equ_gc}II by $1/D$, an order in index $n$ of $\mathcal{O}(1)$ can be proved in the asymptotic behaviour of the limit $n\rightarrow\infty$ for $R_c/R_s<1$. Therefore, the factor $(R_c/R_s)^{(n-2)}$ is the dominant factor in the asymptotic behaviour for the limit $n\rightarrow\infty$ of the {kernel function} of the core $G_{r,n}^\text{(c)}$. Combined with the $n$-dependence of the Legendre polynomials $P_n\left(\cos\theta\right)$ the {general rescaled} radial solution of the core $u_{r}^\text{(c)}/R_c$ at the core radius $R_c$ gives a convergent series at the poles and at the equator, due to the $(R_c/R_s)^n$-dependence ($R_c/R_s<1$, exponential decrease).

The general {rescaled solution} for the radial component of the displacement field for the shell $u_{r}^\text{(s)}/R_s$ is obtained at the outer shell radius $R_s$ as
\begin{align}
\frac{u_{r}^\text{(s)}(R_s\mathbf{e}_r)}{R_s}=\sum_{n=0}^\infty & G^\text{(s)}_{r,n}\left(\frac{\lambda}{E_sR_s},\frac{E_c}{E_s}, \frac{R_c}{R_s},\nu_c,\nu_s\right)\frac{2n+1}{2}P_n\left(0\right)P_n\left(\cos\theta\right)\label{eqn:gs2}
\end{align} 
where {$G^\text{(s)}_{r,n}\left(\lambda/(E_sR_s),E_c/E_s, R_c/R_s,\nu_c,\nu_s\right)$ is the corresponding kernel function of the shell and the remaining factors are the same as for the core solution. In terms of the coefficients $a_n^\text{s}$, $b_n^\text{s}$, $c_n^\text{c}$ and $d_n^\text{c}$, (\ref{eqn:gs2}) can also be written as}
\begin{align}
\frac{u_{r}^\text{(s)}\left(R_s\mathbf{e}_r\right)}{R_s}=\sum_{n=0}^\infty & \left(\frac{a_n^\text{s}}{R_s}R_s^{n+1}(n+1)(-2+n+4\nu_s)+\frac{b_n^\text{s}}{R_s}R_s^{n-1}n\right.\nonumber\\
&\left.+\frac{c_n^\text{s}}{R_s}R_s^{-n}n(3+n-4\nu_s)-\frac{d_n^\text{s}}{R_s}R_2^{-(n+2)}(n+1)\right)P_n\left(\cos \theta\right)\nonumber\\
=\sum_{n=0}^\infty & \frac{2n+1}{2} P_n(0) P_n\left(\cos \theta\right)\frac{\lambda}{E_sR_s}\nonumber\\
&\times \frac{1}{D}\left(\underbrace{\left[\left(\frac{E_c}{E_s}\right)^2\tilde{c}_{05,n}+\left(\frac{E_c}{E_s}\right)\tilde{c}_{06,n}+\tilde{c}_{07,n}\right](n+1)(-2+n+4\nu_s)}_{I}\right.\left.\underbrace{-\left[\left(\frac{E_c}{E_s}\right)^2\tilde{c}_{08,n}+\left(\frac{E_c}{E_s}\right)\tilde{c}_{09,n}+\tilde{c}_{10,n}\right]n}_{II}\right.\nonumber\\
&\left.\underbrace{+\left(\frac{R_c}{R_s}\right)^{(2n-1)}\left[\left(\frac{E_c}{E_s}\right)^2\tilde{c}_{11,n}+\left(\frac{E_c}{E_s}\right)\tilde{c}_{12,n}+\tilde{c}_{13,n}\right]n(3+n-4\nu_s)}_{III}\right.\left.\underbrace{+\left(\frac{R_c}{R_s}\right)^{(2n+1)}\left[\left(\frac{E_c}{E_s}\right)^2\tilde{c}_{14,n}+\left(\frac{E_c}{E_s}\right)\tilde{c}_{15,n}+\tilde{c}_{16,n}\right](n+1)}_{IV}\right)\label{equ_gs}\nonumber\\
\end{align}
\end{@twocolumnfalse}
]
\twocolumn[
\begin{@twocolumnfalse}
By comparing the solution for $u_{r}^\text{(s)}/R_s$ with that in (\ref{eqn:gs2}), it can be concluded that the {kernel function} of the shell $G_{r,n}^\text{(s)}$ is the product of the factors $\lambda/(E_sR_s)$, $1/D$ and the sum of \ref{equ_gs}I + \ref{equ_gs}II + \ref{equ_gs}III + \ref{equ_gs}IV. By multiplying the sum \ref{equ_gs}I + \ref{equ_gs}II by $1/D$, an order in index $n$ of $\mathcal{O}(1/n)$ can be proved in the asymptotic behaviour of the limit $n\rightarrow\infty$ for $R_c/R_s<1$. Multiplying the sum \ref{equ_gs}III + \ref{equ_gs}IV by $1/D$ leads to a dominant factor of $(R_c/R_s)^{2n}$ under the same conditions. Therefore, the asymptotic behaviour for $n\rightarrow\infty$ is proportional to $1/n$ for the {kernel function} of the shell $G_{r,n}^\text{(s)}$. Combined with the $n$-dependence of the Legendre polynomials $P_n\left(\cos\theta\right)$ the {general rescaled} radial solution for the shell $u_{r}^\text{(s)}/R_s$ at the outer shell radius $R_s$ results in a divergent series at the equator $(\theta=\pi/2)$, due to the $1/n$-dependence of $G_{r,n}^\text{(s)}$ (harmonic series) and a convergent series at the poles $(\theta=0, \pi)$, due to the property of the Legendre polynomials at the poles (alternating series and a monotonic decrease to zero of the absolute value of the summands).\\
\end{@twocolumnfalse}
]
\pagebreak
\bibliography{acs-manuscript_core_shell.bib}
\bibliographystyle{rsc} 
\end{document}